\documentclass[pre,twocolumn,aps,showpacs]{revtex4}
\usepackage{graphics,graphicx,epsfig}
\usepackage{color}
\begin{document}
\title{Crossover from anomalous to normal diffusion in porous media}
\author{F. D. A. Aar\~ao Reis${}^{1,}$\footnote{Email address: reis@if.uff.br}\\
Dung di Caprio${}^{2,}$\footnote{Email address: dung.dicaprio@yahoo.fr}
}
\affiliation{
(1) Instituto de F\'\i sica, Universidade Federal Fluminense,\\
Avenida Litor\^anea s/n, 24210-340 Niter\'oi RJ, Brazil,\\
(2) Institut de Recherche de Chimie Paris, CNRS - Chimie ParisTech,
11, rue P. et M. Curie, 75005 Paris, France}
\date{\today}

\begin{abstract}
Random walks (RW) of particles adsorbed in the internal walls of porous deposits produced
by ballistic-type growth models are studied. The particles start at the external surface of the
deposits and enter their pores, in order to simulate an external flux of a species towards a porous
solid. For short times, the walker concentration decays as a stretched exponential of the depth $z$,
but a crossover to long time normal diffusion is observed in most samples.
The anomalous concentration profile remains at long times in very porous solids if the walker
steps are restricted to nearest neighbors and is accompanied with subdiffusion features.
These findings are correlated with a decay of the explored area with $z$.
The study of RW of tracer particles left at the
internal part of the solid rules out an interpretation by diffusion
equations with position-dependent coefficients.
A model of RW in a tube of decreasing cross section explains those
results by showing long crossovers from an effective subdiffusion regime to an asymptotic
normal diffusion. The crossover position and density are analytically calculated for a tube with area
decreasing exponentially with $z$ and show good agreement with numerical data.
The anomalous decay of the concentration profile is interpreted as a templating effect of
the tube shape on the total number of diffusing particles at each depth,
while the volumetric concentration in the actually explored
porous region may not have significant decay.
These results may explain the anomalous diffusion of metal atoms in porous deposits
observed in recent works.
They also confirm the difficulty in interpreting experimental or computational data on
anomalous transport reported in recent works, particularly if only the concentration profiles
are measured.

\end{abstract}

\maketitle

\section{Introduction}
\label{intro}

Diffusion in porous media long has been a topic of interest \cite{bouchaud,havlin}
due to a variety of technological applications, which are usually related to the large
surface-to-volume ratio of those materials.
The concentration profile of a diffusing species entering a disordered medium is
typically of the form
\begin{equation}
\rho \left( z,t\right) \sim \exp{\left[ -z^\mu/R\left( t\right)\right] } ,
\label{rho}
\end{equation}
where $z$ is the direction perpendicular to the external surface, with $z=0$ at that
surface. The function $R\left( t\right)$ measures the spreading of the concentration
profile in time $t$ and asymptotically behaves as
\begin{equation}
R\left( t\right) \sim t^\alpha .
\label{R}
\end{equation}
From Eqs. (\ref{rho}) and (\ref{R}), the mean-square displacement of the diffusing material scales as
\begin{equation}
\langle z^2\rangle \sim t^{2\alpha/\mu} .
\label{z2}
\end{equation}
Normal (Fickean) diffusion is characterized by a spatially Gaussian
distribution of the diffusing species and linearly increasing mean-square displacement:
$\mu=2$ and $\alpha=1$. Otherwise, the diffusion is called anomalous. The scaling of $\langle z^2\rangle$
is frequently used to separate cases of subdiffusion ($\alpha/\mu<1/2$) and superdiffusion ($\alpha/\mu>1/2$).
The most frequently observed anomaly is subdiffusion because the irregularities of
the porous media (e. g. barriers and dead ends) restrict the movement of the diffusing species.

The relevance of anomalous diffusion in real systems has been shown in several recent works.
For instance, fluorescence spectroscopy was used to distinguish cases of normal
and subdiffusion of tracer particles in colloidal crystals \cite{raccis}, and combined with
confocal microscopy (to probe concentration profiles), it was used to show that reaction-diffusion
models were necessary to explain dye transport in metal-organic frameworks \cite{han2012}.
Nuclear magnetic resonance was recently used to measure exponents $\alpha$ and $\mu$
of water diffusion in several disordered colloidal systems, showing deviations from 
Fickean diffusion \cite{palombo2011,palombo2013}.
Particularly interesting cases of stretched exponential concentration profiles and superdiffusion
were shown when $Pt$ atoms entered the pores of porous carbon and anodic porous alumina during plasma sputtering
deposition on those samples \cite{brault1,brault2,brault3}. Moreover, due to the importance of
subdiffusion phenomena in cells and model systems, the need for experimental standards was recently
highlighted \cite{saxton}. 

There is also an intense theoretical work on anomalous diffusion.
Some approaches involve the computational study of fluid flow and solute transport in disordered media
\cite{bijeljic,deanna,moore,roubinet,sahimi,vasilyev},
analytical methods based on advection-diffusion equations
\cite{neuman,wang} and fractional diffusion equations \cite{metzler,lenzi2011,malacarne,balankin2012a}.
The study of random walks (RW) in models of disordered media \cite{bouchaud,havlin} is another simple and widely
used approach, whose relevance to describe
real system features is illustrated in recent works \cite{korabel,novikov,giri}. 

Motivated by the observation of anomalous diffusion of $Pt$ atoms penetrating samples of disordered
porous carbon \cite{brault1}, in the first part of this work we study
RW of particles adsorbed in the internal surface of porous deposits after being
released at their outer surface. The deposits are produced by ballistic deposition (BD) \cite{fv,vold}
and by an extension of that model \cite{hivert,katzavbd,sigma,khanin}. For short times,
the concentration profiles are
stretched exponentials [slower than Gaussian, $\mu<2$ in Eq. (\ref{rho})]. This
feature is still observed at long times in one type of deposit, and the corresponding time scaling
of the concentration profile is consistent with subdiffusion.
The area explored by the walkers decreased approximately as the inverse of the depth in this case.
Tracer particles inside the samples do not show dependence of diffusion coefficients
on the depth, which rules out theoretical approaches with this assumption.
Although the values of exponents $\mu$ and $\alpha$ differ from those of Ref. \protect\cite{brault1},
the scaling of the concentration with depth and time show similar deviations from normal diffusion.

The second part of this work is devoted to the study of random walks in tubes of decreasing cross section,
which show concentration profiles with stretched exponentials and anomalous scaling of the
mean-square displacement. A tube with area decreasing with the inverse of the depth (similarly to
the porous deposits with anomalous diffusion) shows significant deviations from normal scaling and
exponent values much closer to those of $Pt$ atoms entering porous carbon samples \cite{brault1}.
A detailed discussion of the crossover from anomalous to normal diffusion is presented for the
case of tubes with exponentially decreasing cross section, in which analytical solution is possible.
These results confirm the difficulty in interpreting experimental or computational data on anomalous
transport, particularly the concentration profiles, consistently with recent works
\cite{saxton,raccis,korabel,kennedy}.
 
This paper is organized as follows. In Sec. II, we present the models of
ballistic deposits and the random walk models.
In Sec. III, we study the scaling of concentration profiles inside the ballistic deposits
and related quantities.
In Sec. IV, we study diffusion in tubes of decreasing cross section and discuss the crossover
in concentration scaling.
In Sec. V, our conclusions are presented.

\section{Porous deposits and random walk simulation}
\label{models}

Here we describe the two types of disordered porous media used in this work and the
two types of RW simulated inside those samples.

In the BD model, particles are
released from a randomly chosen position above a $d$-dimensional substrate, follow
trajectories perpendicular to an initially flat substrate [here a $(x,y)$ plane] and stick upon the first
contact with a nearest neighbor (NN) occupied site, which may be the substrate or a previously
deposited particle \cite{fv,vold}. The aggregation rules are illustrated in Fig. 1a.
The resulting aggregate is porous and has a rough surface, as illustrated in the
cross-sectional view of a three-dimensional deposit in Fig. 1b.
Long time simulations indicate that the porosity is approximately $0.667$.

\begin{figure}
\includegraphics[width=3cm]{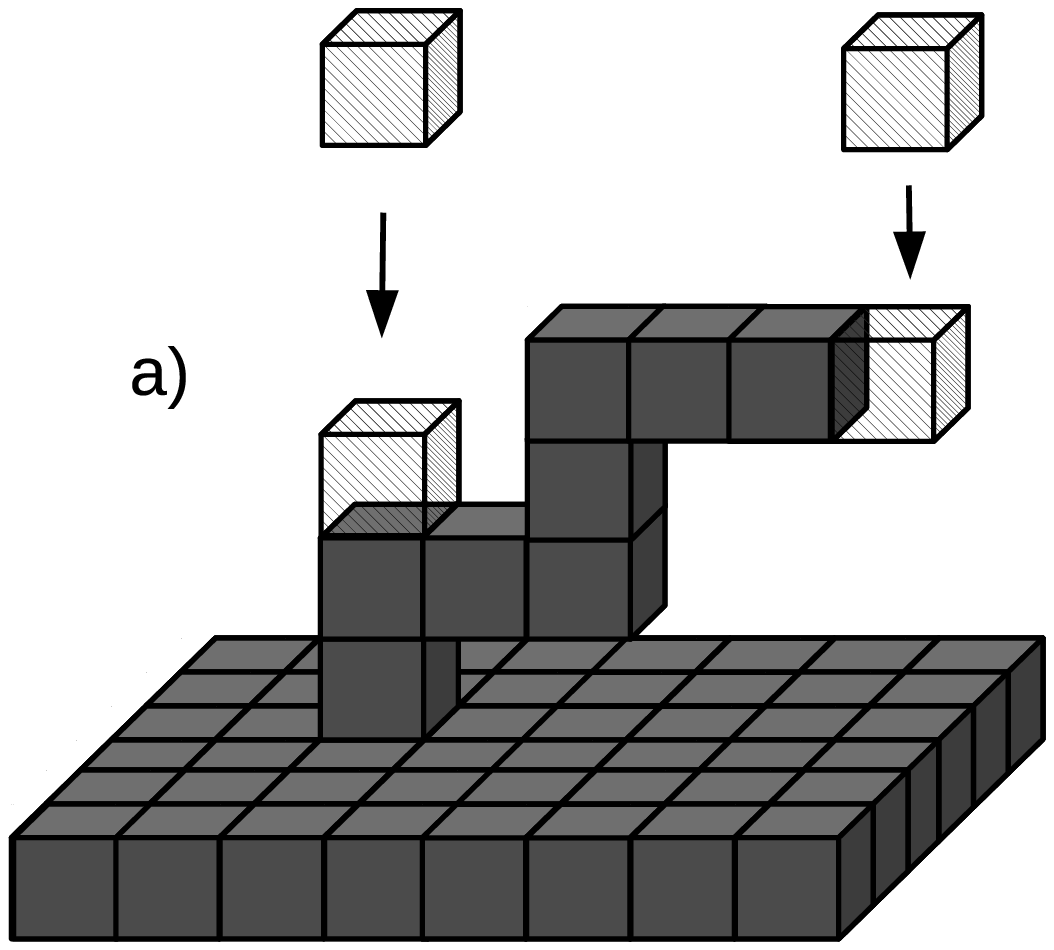}
\hspace{2ex}
\includegraphics[width=3cm]{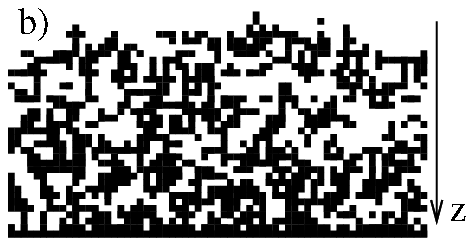}\\[2ex]
\includegraphics[width=3cm]{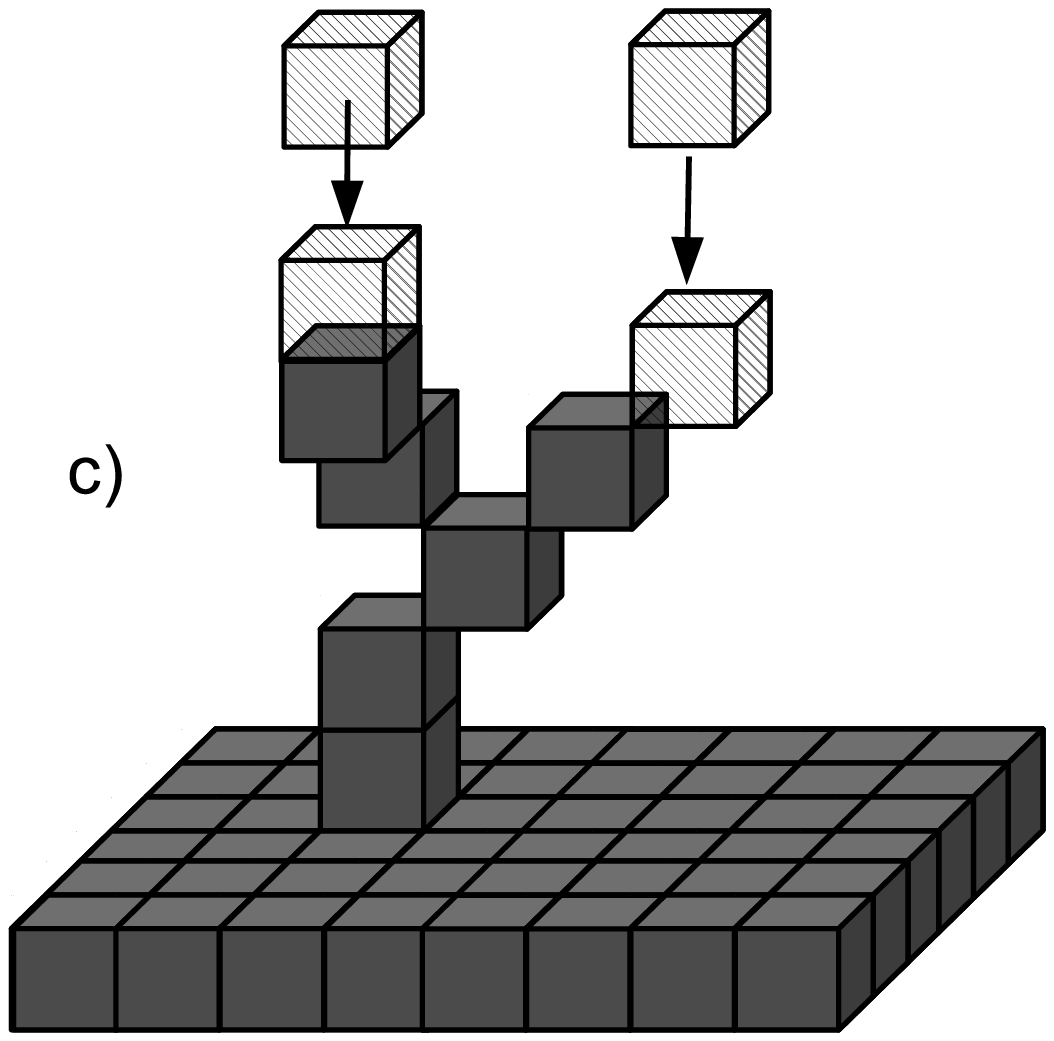}
\hspace{2ex}
\includegraphics[width=3cm]{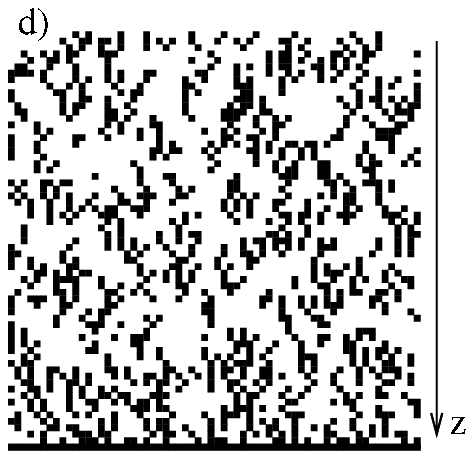}
\caption{(Color online) Illustration of the aggregation rules and cross sections of the tops of deposits
generated by (a),(b) BD and (c),(d) BDNNN. Dark gray cubes are previously aggregated solid particles
and light gray cubes are incident particles, whose aggregation positions (also shown in light gray) are
indicated by arrows. 
}
\label{fig1}
\end{figure}

A simple extension of that model, called BDNNN, allows aggregation of the incident particle
by contact with a NN or a next nearest neighbor (NNN) occupied site \cite{hivert,katzavbd,sigma,khanin},
as illustrated in Fig. 1c.
The resulting deposit has a larger porosity, approximately $0.834$,
as illustrated by the cross-sectional view of Fig. 1d with an identical number of
deposited layers as in Fig. 1c.

Deposits of average heights near $900$ lattice units and lateral sizes $1024$
were grown. The structure of each sample remains fixed during the simulation of diffusion, i.e.
they are nondeformable porous solids. A single deposit is large enough to
represent all the microscopic environments that are relevant for the RW statistics.
For this reason, only $3$ samples were produced by each growth model (BD and BDNNN)
and used in RW simulations, providing approximately the same average quantities;
for instance, the dispersion in the porosity is smaller than $0.3\%$.

The maximal height of a solid particle in each column $(x,y)$ of a deposit is defined
as the column height $h(x,y)$. The set $\{ h(x,y)\}$ defines the external surface of the
deposit. The average $\overline{h}$ of that set is taken as the position $z=0$ for RW simulations.
The $z$ axis is oriented to the interior of the deposit, perpendicularly
to the substrate where it was grown. This is illustrated in Figs. 1b and 1d.

In each porous sample, $2^{25}$ ($\sim {10}^8$) non-interacting walkers are left at $t=0$
at randomly chosen points with $z=0$, with the condition that at least one NN of the starting
point is a solid site.
In one time unit, one step trial is performed for each walker.
Adsorption to the porous solid is always required, i. e. only steps to points that
also have a NN solid site are allowed.

Independent simulations were performed with two possible conditions for choosing the steps of the
walkers: (I) a step to a NN site of the underlying cubic lattice (6 possibilities)
is randomly chosen and (II) a step to a NN or a NNN site of the underlying
cubic lattice (18 possibilities) is randomly chosen. If the target site has at least one NN of the porous solid
(adsorption condition), then the step is executed. Otherwise, the step trial is rejected and
the walker remains at the same position.

Figs. 2a and 2b illustrate steps with conditions (I) and (II), respectively. Diffusion under condition
(I) is severely restricted because it does not allow corner rounding. However, with condition (II),
this process is possible and steps from one branch of the porous solid to another one are facilitated.

\begin{figure}
\includegraphics[width=3cm]{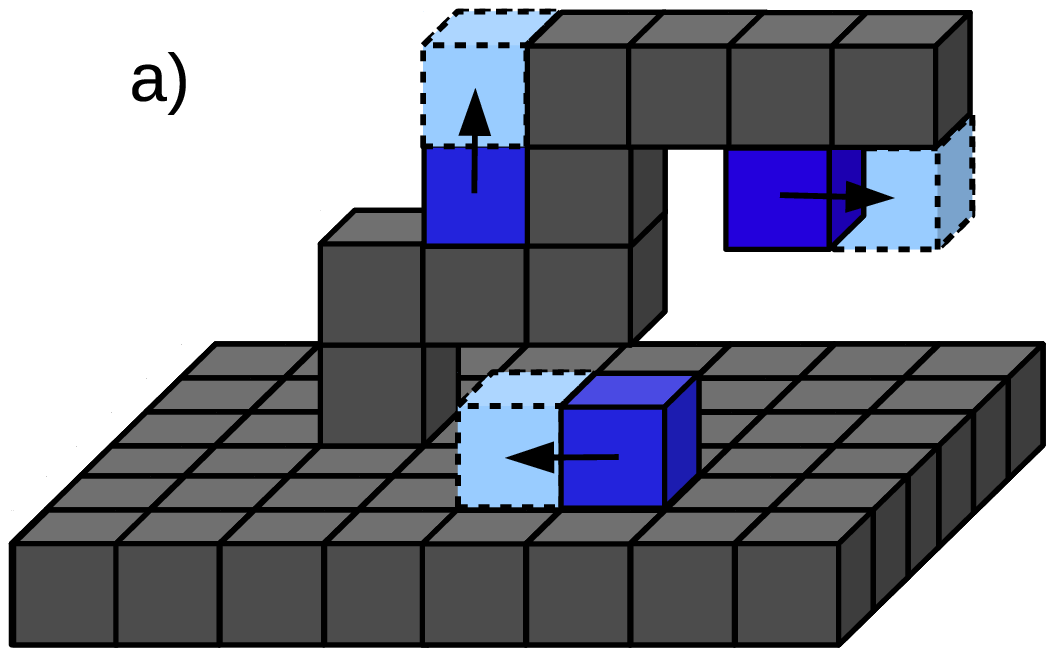}
\hspace{2ex}
\includegraphics[width=3cm]{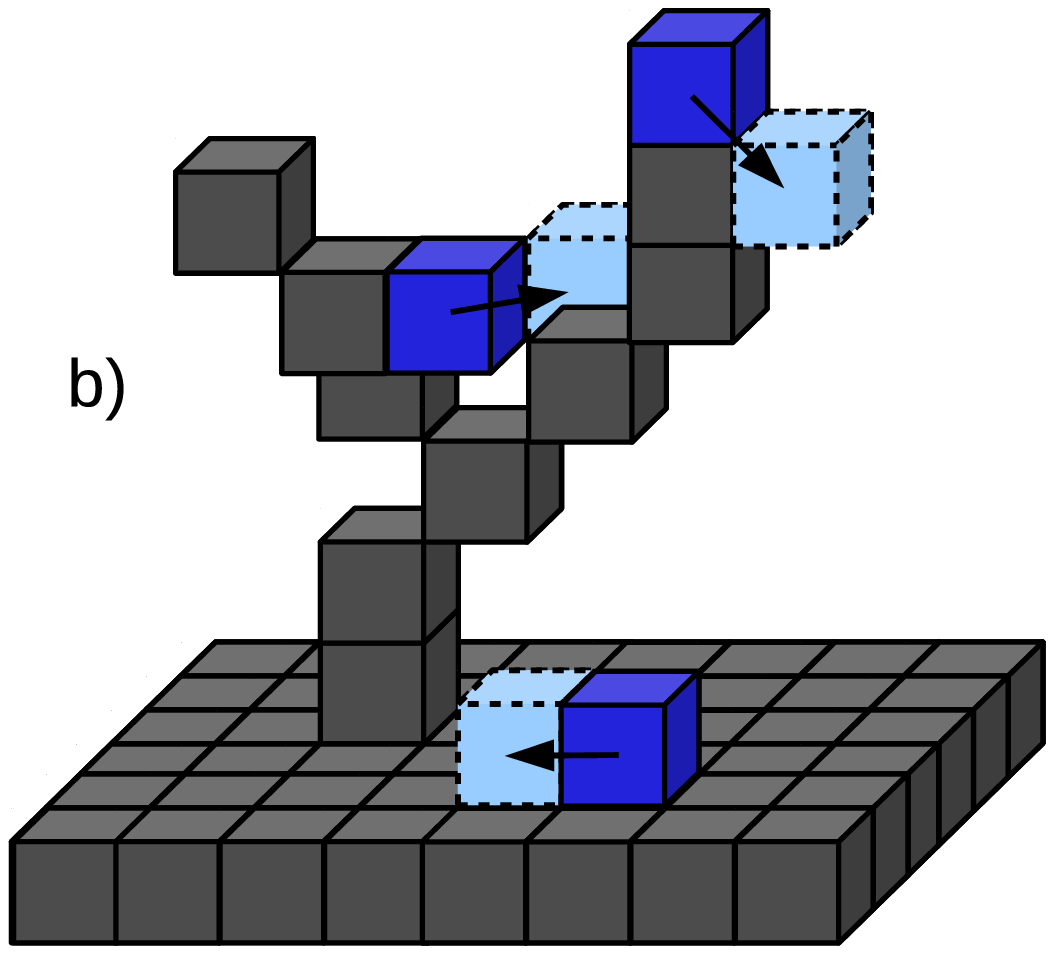}
\caption{(Color online) Illustration of possible steps of RW of the species (dark blue) moving inside the
porous medium (solid part in dark gray) according to rules (a) (I) and (b) (II). The target position of
the allowed steps is shown in light blue.
}
\label{fig2}
\end{figure}

The maximum simulation time is ${10}^5$, corresponding to the maximum number of possible steps
of each walker (due to the no-desorption condition, the number of executed steps may be much
smaller than that). This maximal time is suitable to avoid that any walker reaches the bottom of the substrate.

As a final remark, we recall that there are other extensions of BD to represent a variety of
porous materials \cite{bbd,perez,bbdflavio,supti2007,robledo,juvenil}.
However, the aim of this work is to understand basic features of anomalous diffusion of
a species entering a porous material, instead of representing a particular application.
For this reason, we restrict our work to the study of samples produced by BD and BDNNN models.

\section{Diffusion in the porous deposits}
\label{porous}

\subsection{Scaling of concentration profiles}
\label{concentration}

Figs. 3a and 4a show $\log{\left[ -\log{(\rho/\rho_0)}\right]}$ as a function of $\log{z}$
in three different times for RW in porous solids produced by BD (lower porosity - Fig. 1b),
respectively with conditions (I) and (II) for the steps. Here, $\rho_0$ is the maximum value of
the density profile. 
Figs. 5a and 6a show the same quantities for RW in porous solids produced by BDNNN
(high porosity - Fig. 1d), respectively with conditions (I) and (II) for the steps.

\begin{figure}
\includegraphics[width=6cm]{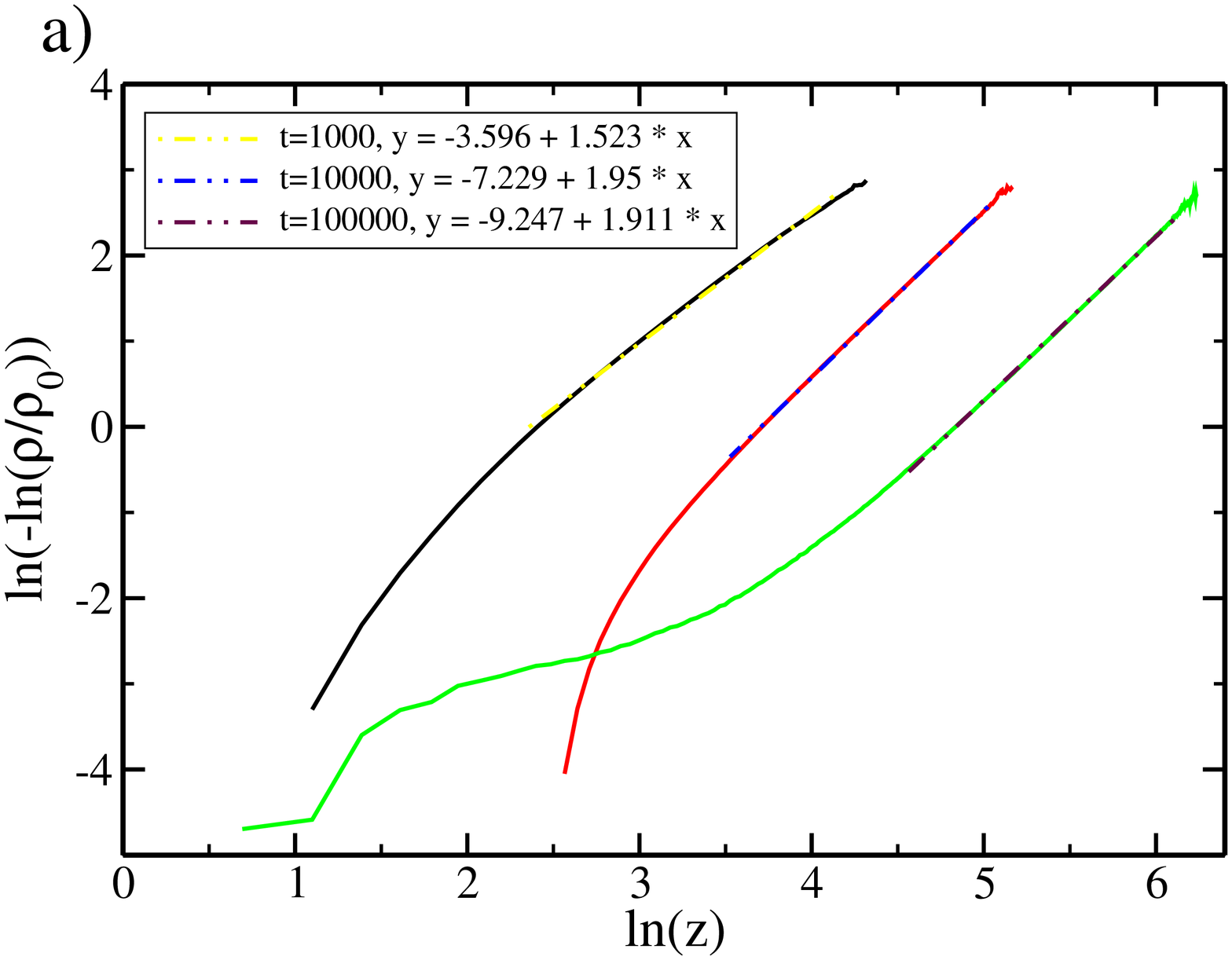}\\[2ex]
\includegraphics[width=6cm]{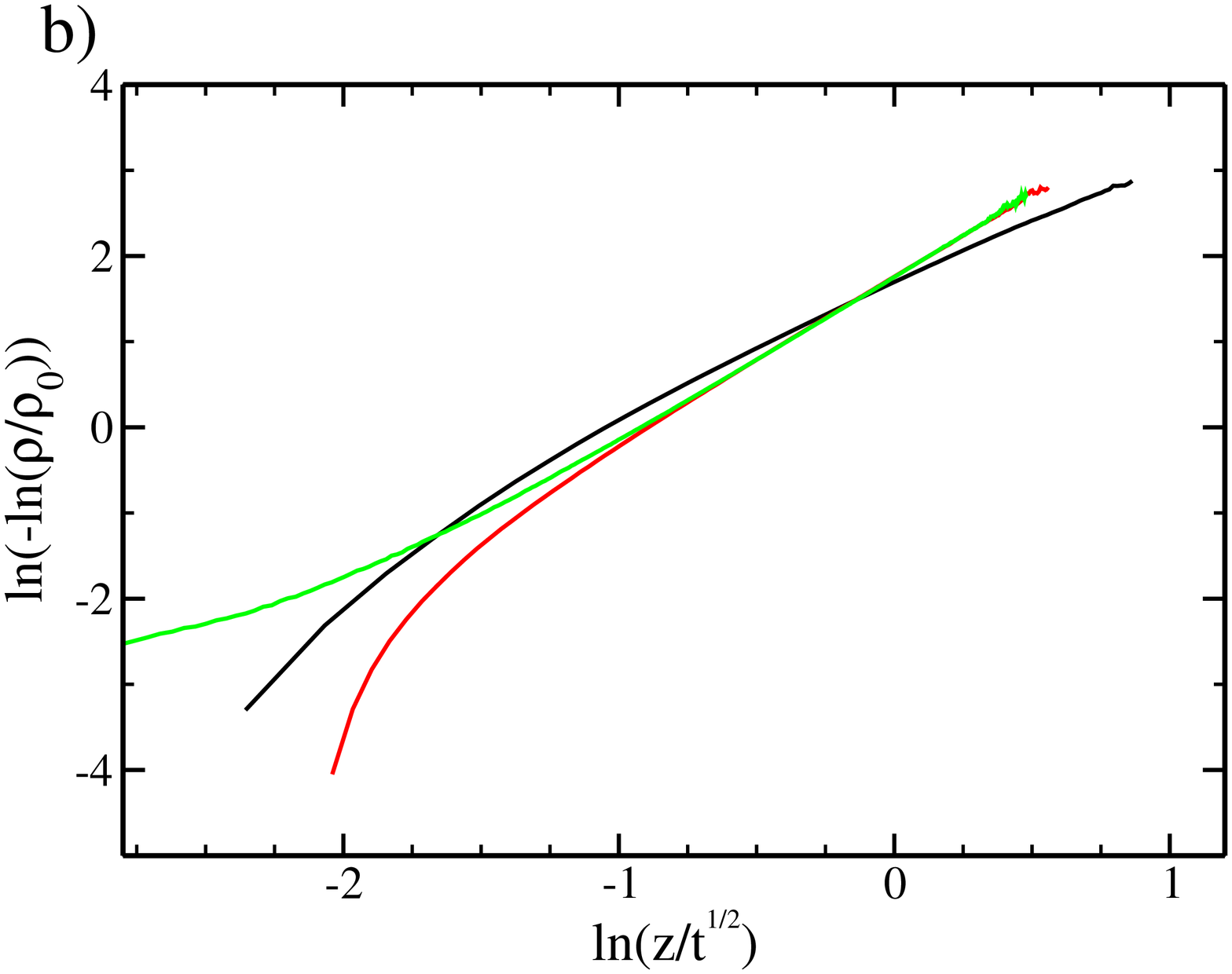}\\[2ex]
\includegraphics[width=6cm]{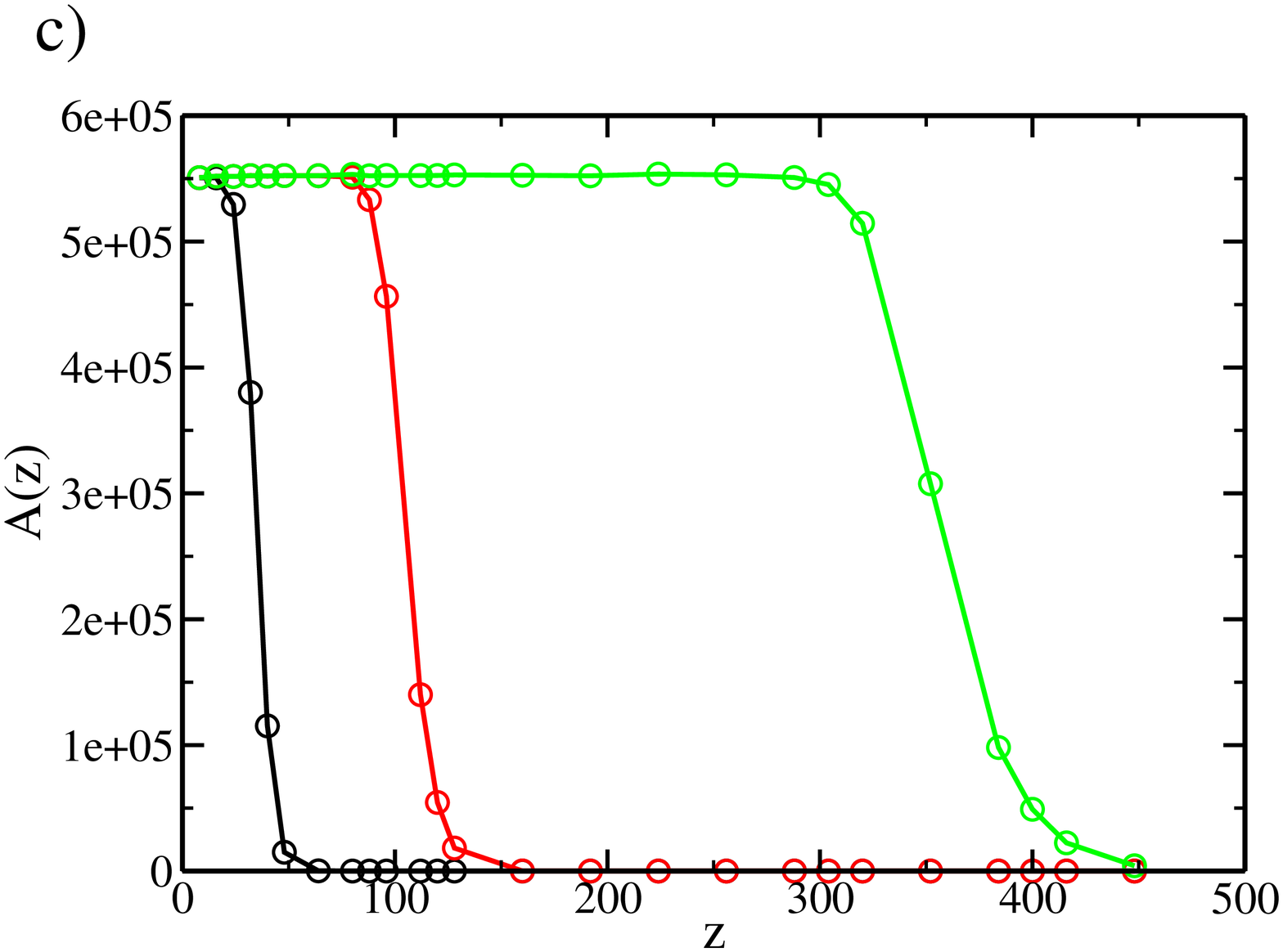}\\[2ex]
\includegraphics[width=6cm]{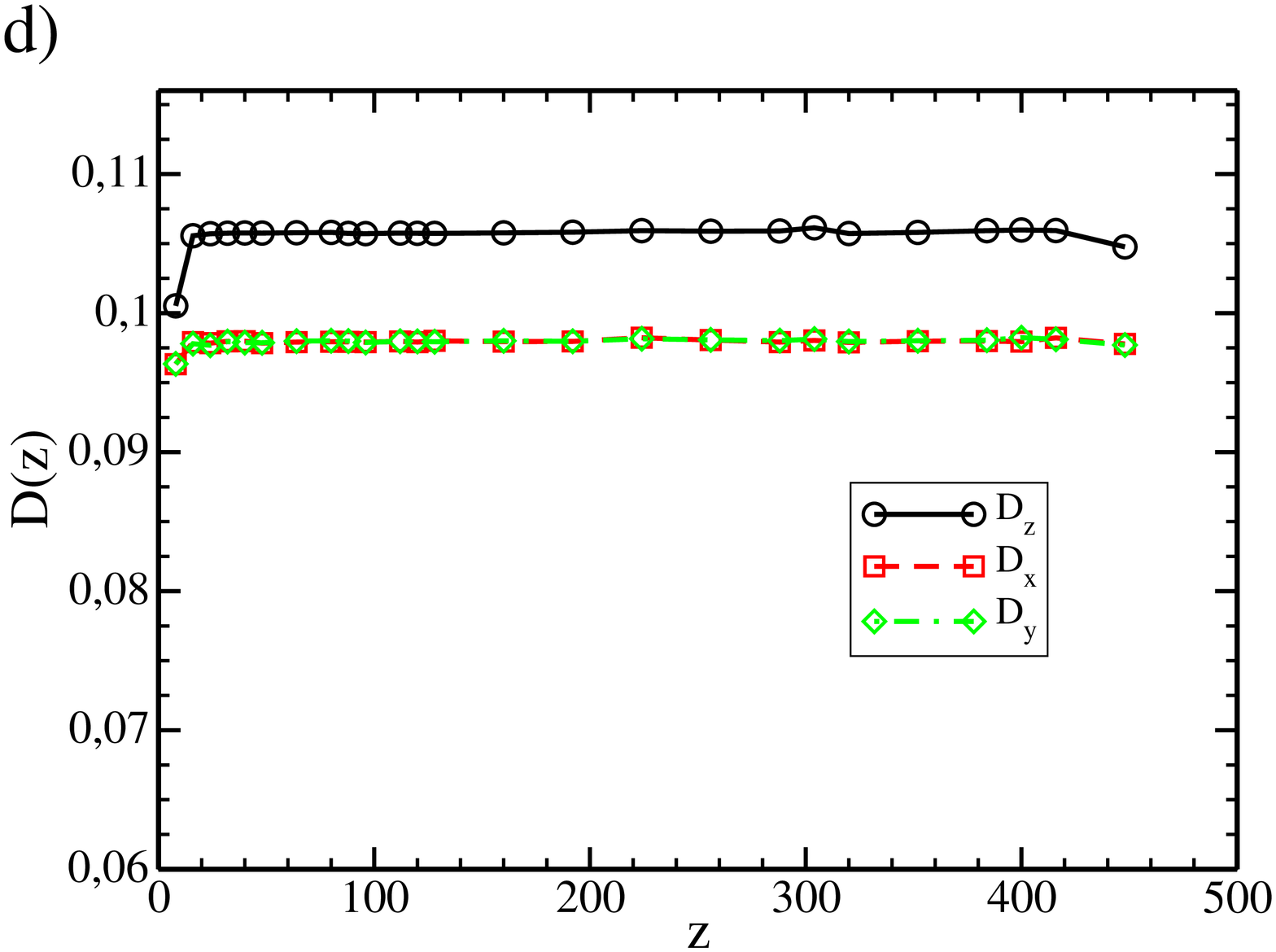}
\caption{(Color online)
Concentration profiles without (a) and with (b) time scaling, number of sites explored by RW (c), and
local diffusion coefficients (d) of tracer particles as a function of the depth $z$
in porous solids produced by BD with rule (I) for the random steps of the RW.
In (a), (b) and (c), red squares, green crosses and blue triangles relate respectively to $t=1000, 10000, 100000$.
Full, dashed, and dash-dot-dot lines in (a) are linear fits for the largest values of $z$.
}
\label{fig3}
\end{figure}

\begin{figure}
\includegraphics[width=6cm]{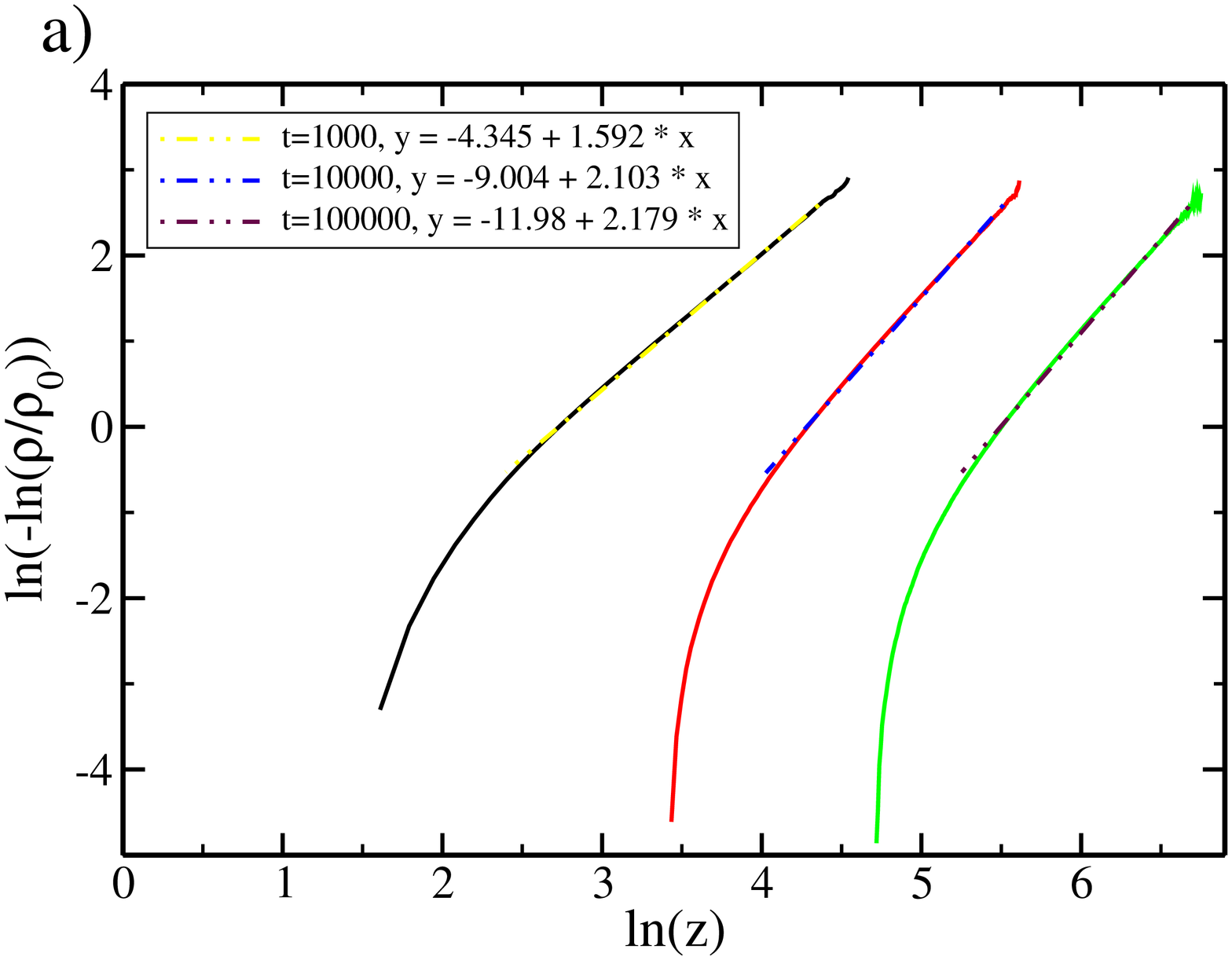}\\[2ex]
\includegraphics[width=6cm]{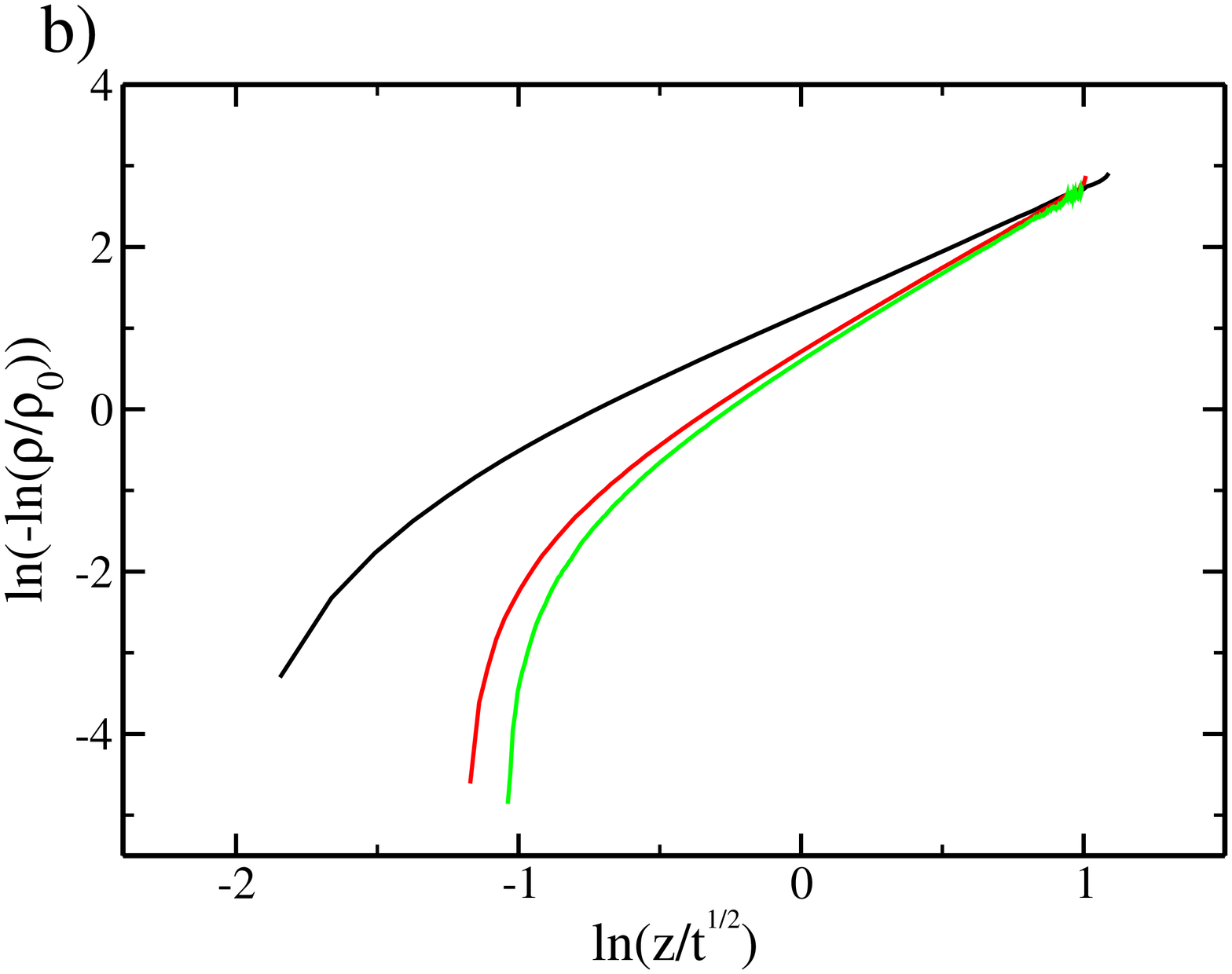}\\[2ex]
\includegraphics[width=6cm]{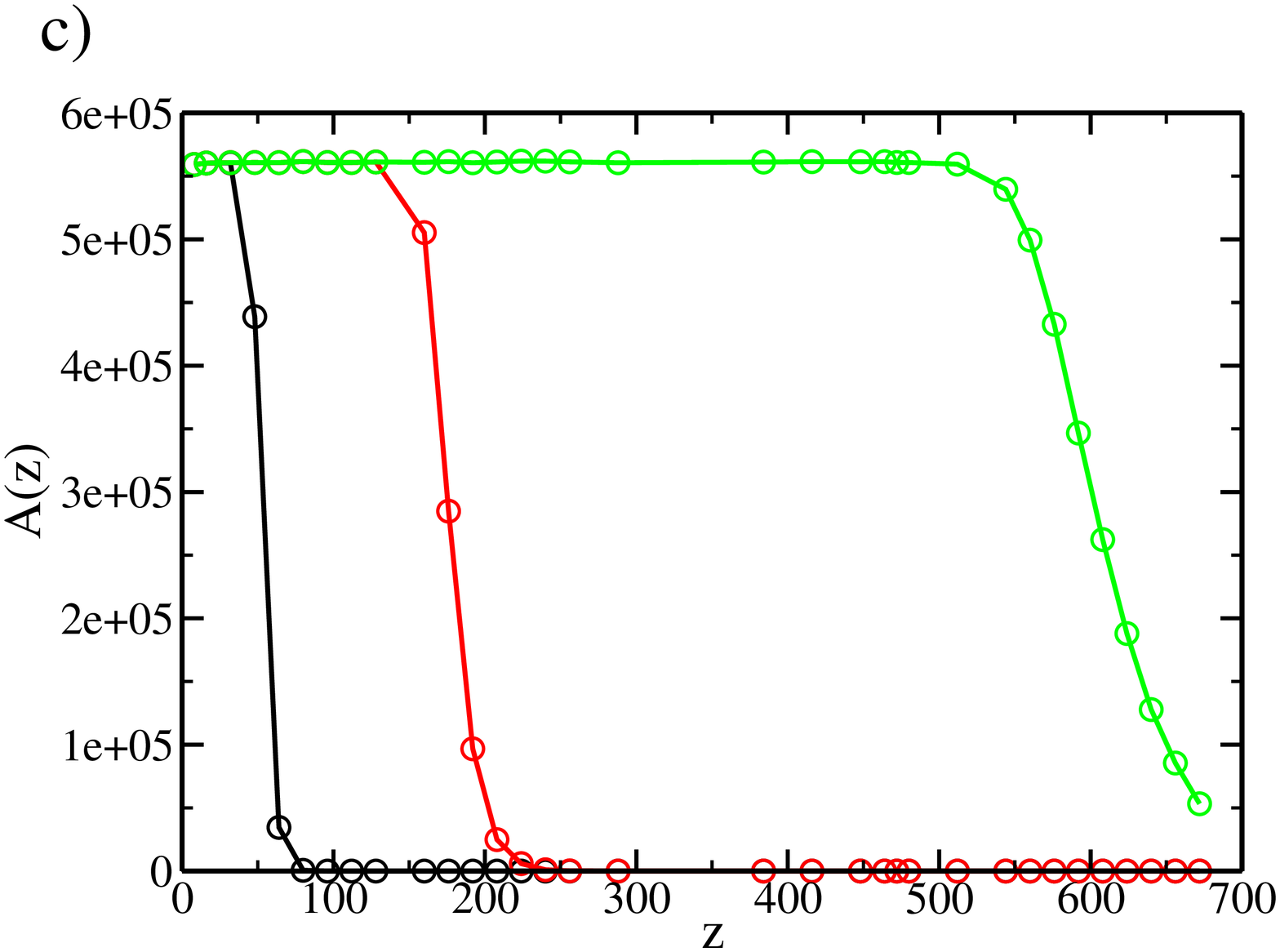}\\[2ex]
\includegraphics[width=6cm]{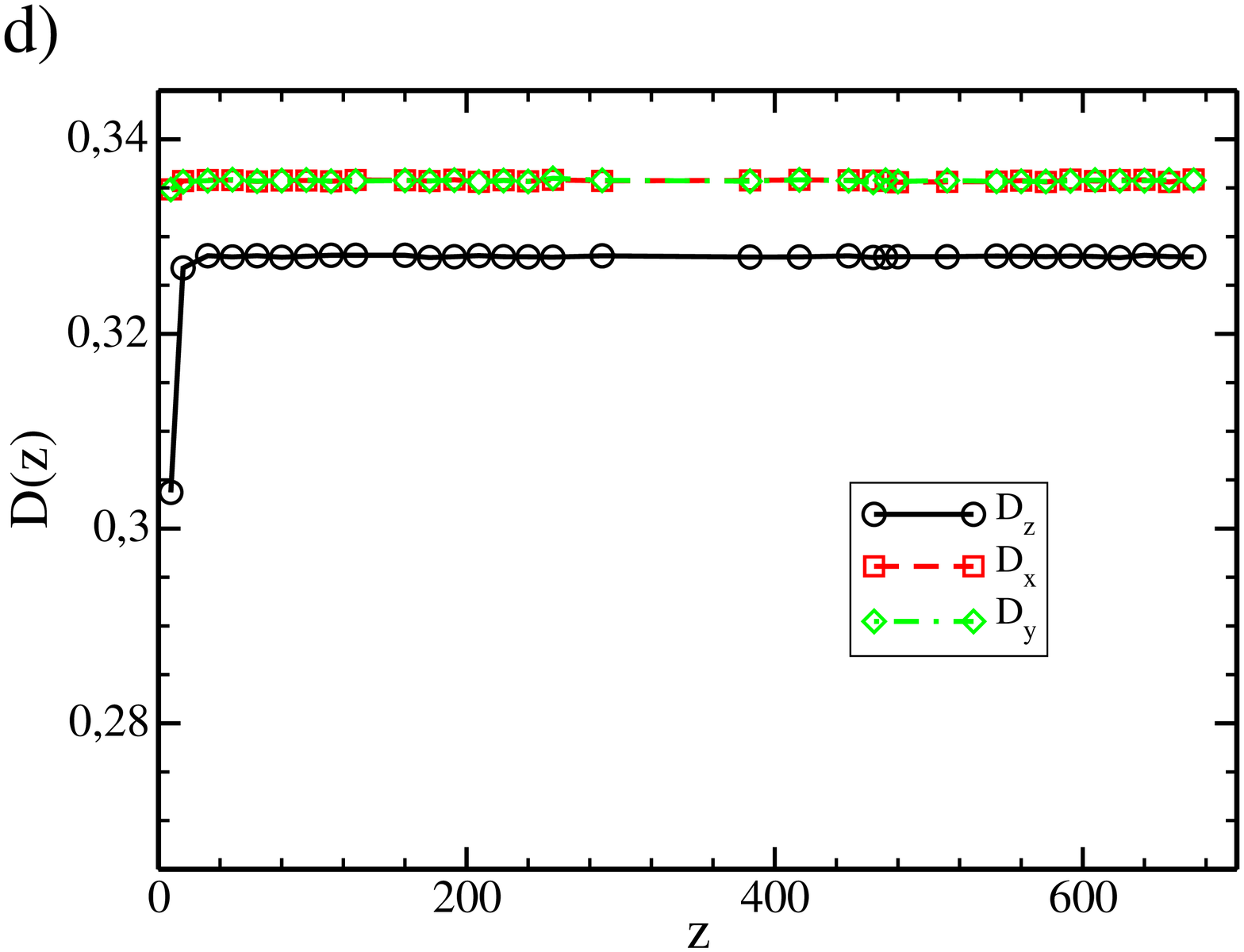}
\caption{(Color online)
Concentration profiles without (a) and with (b) time scaling, number of sites explored by RW (c), and
local diffusion coefficients (d) of tracer particles as a function of the depth $z$
in porous solids produced by BD with rule (II) for the random steps of the RW.
Points and lines are defined in the caption to Fig. \ref{fig3}.
}
\label{fig4}
\end{figure}

\begin{figure}
\includegraphics[width=6cm]{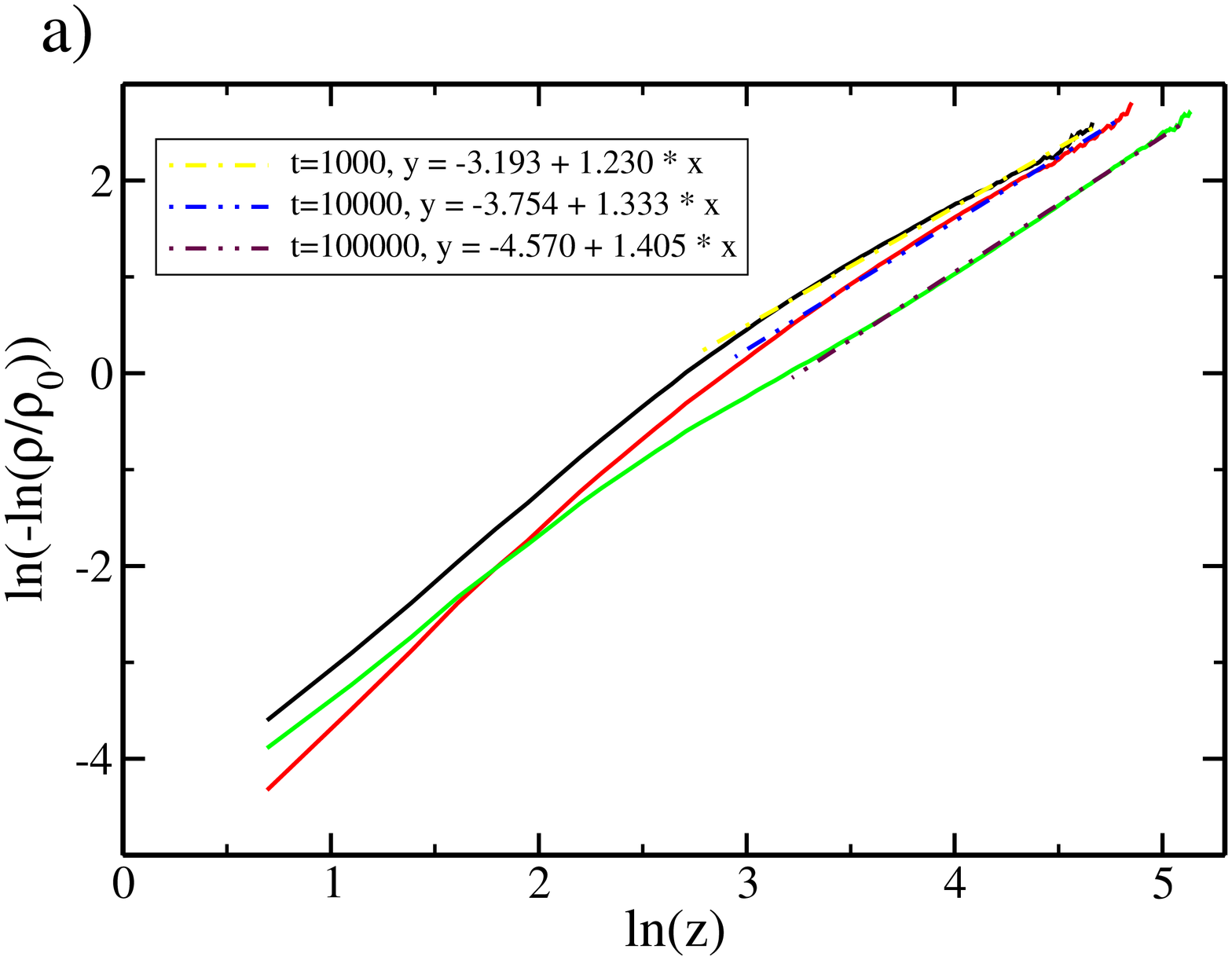}\\[2ex]
\includegraphics[width=6cm]{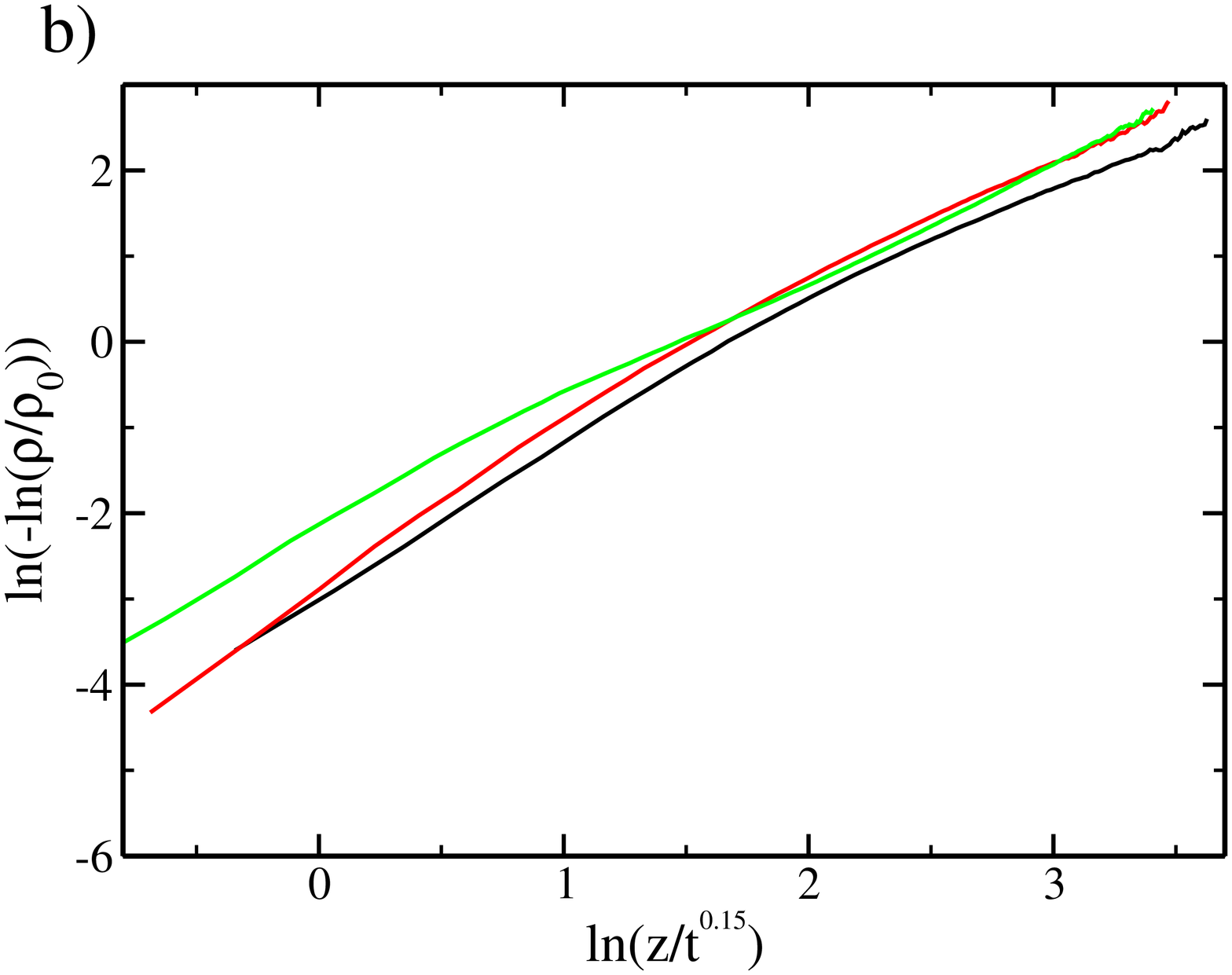}\\[2ex]
\includegraphics[width=6cm]{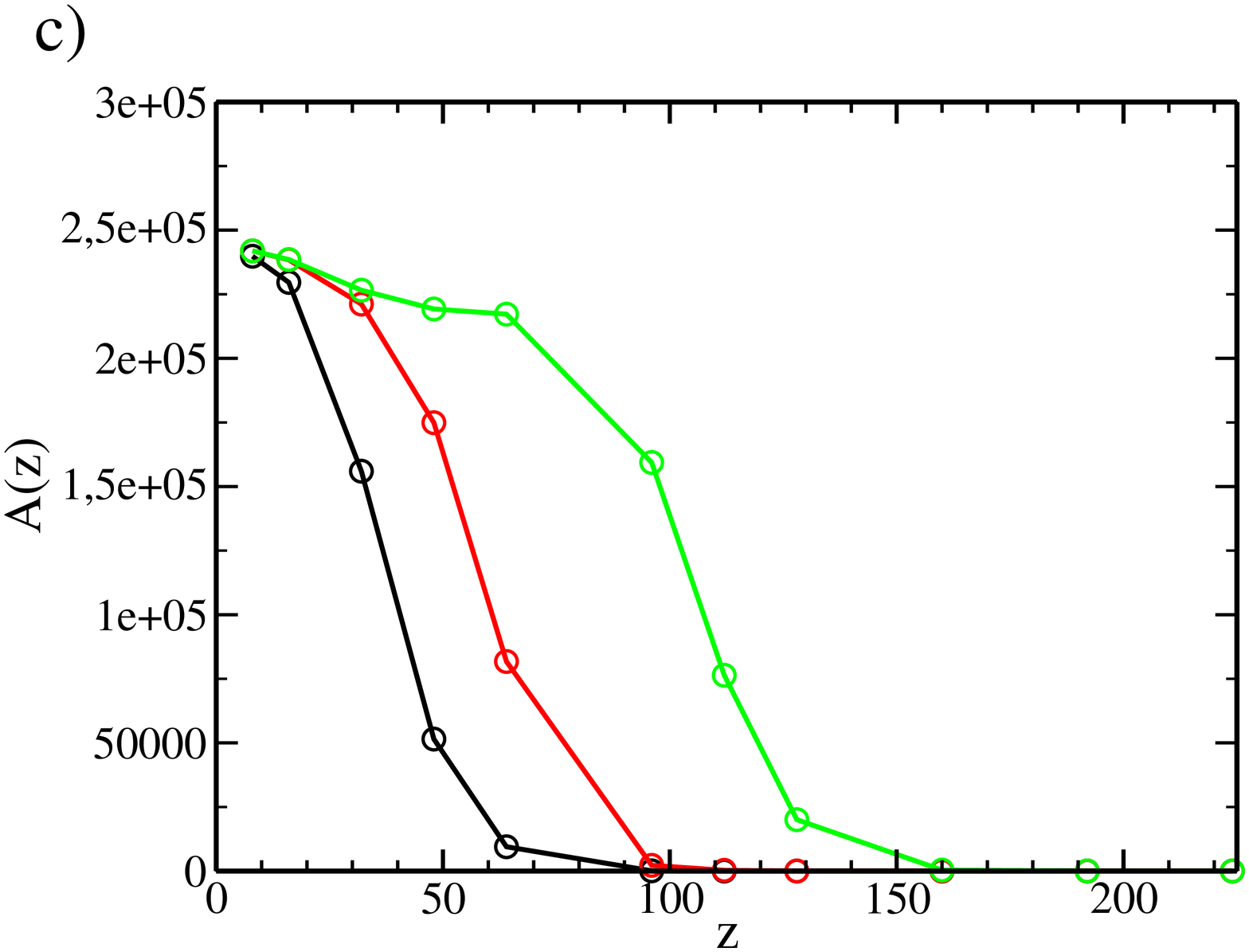}\\[2ex]
\includegraphics[width=6cm]{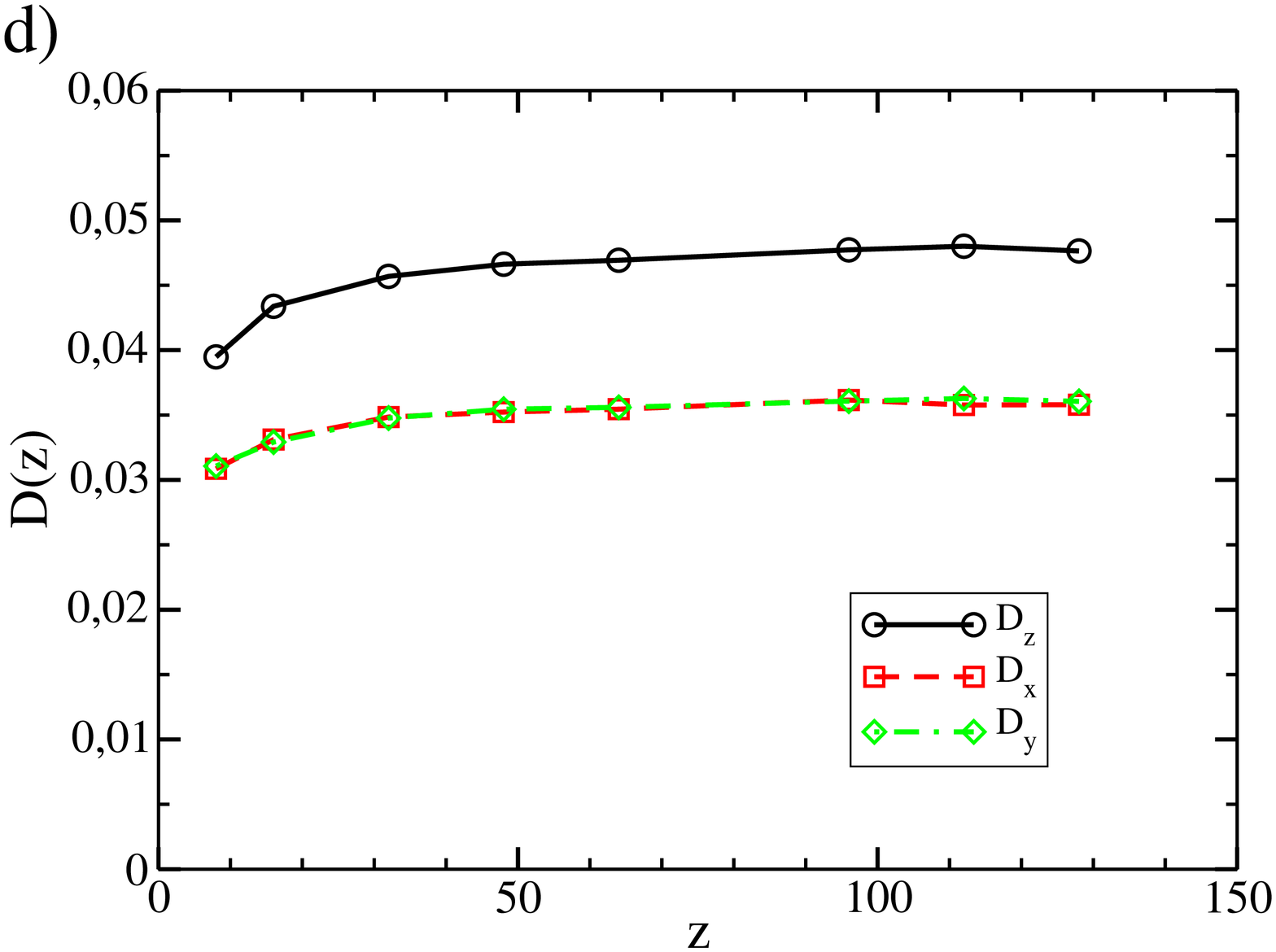}
\caption{(Color online)
Concentration profiles without (a) and with (b) time scaling, number of sites explored by RW (c), and
local diffusion coefficients (d) of tracer particles as a function of the depth $z$
in porous solids produced by BDNNN with rule (I) for the random steps of the RW.
Points and lines have identical meaning as in Figure \ref{fig3}.
}
\label{fig5}
\end{figure}

\begin{figure}
\includegraphics[width=6cm]{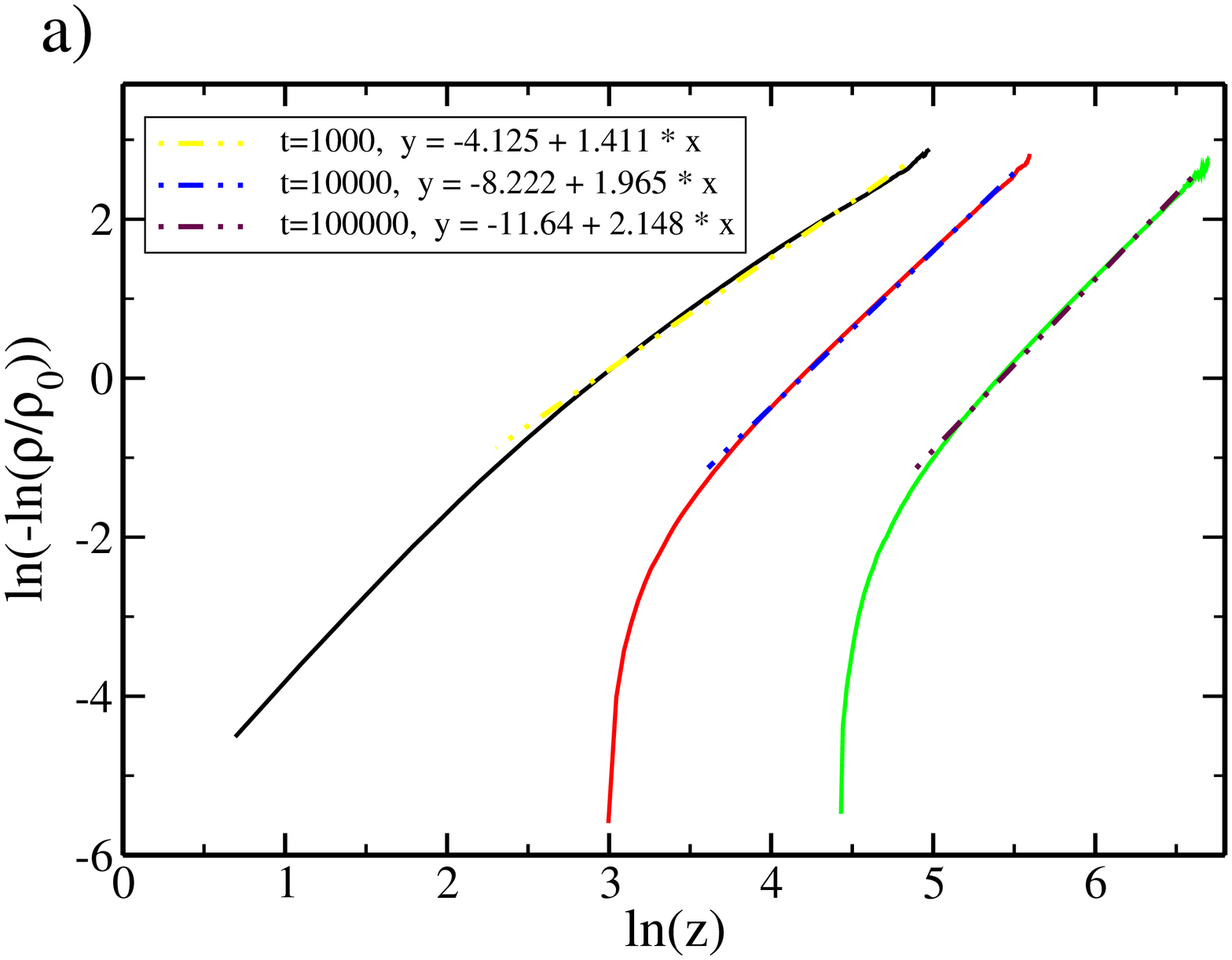}\\[2ex]
\includegraphics[width=6cm]{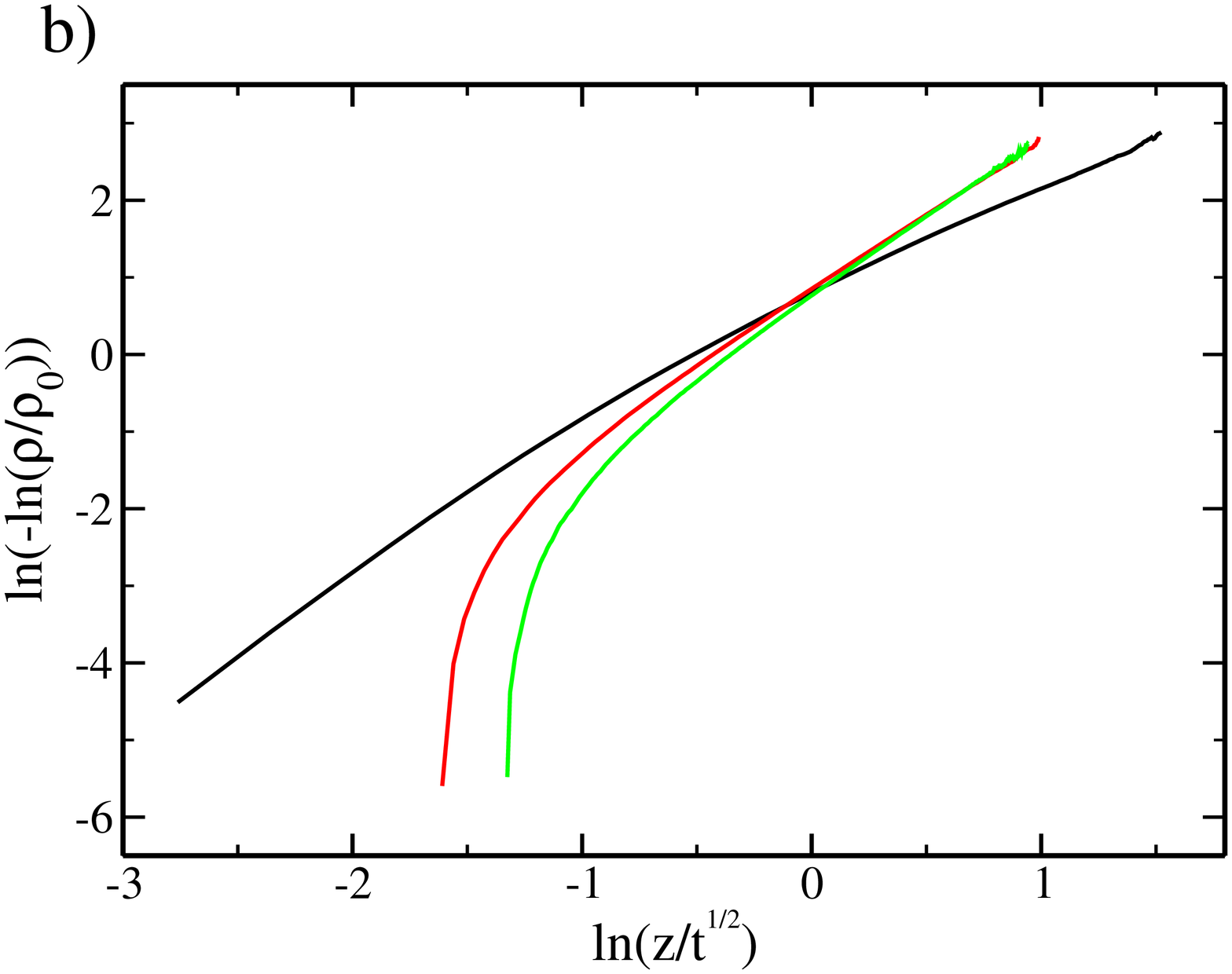}\\[2ex]
\includegraphics[width=6cm]{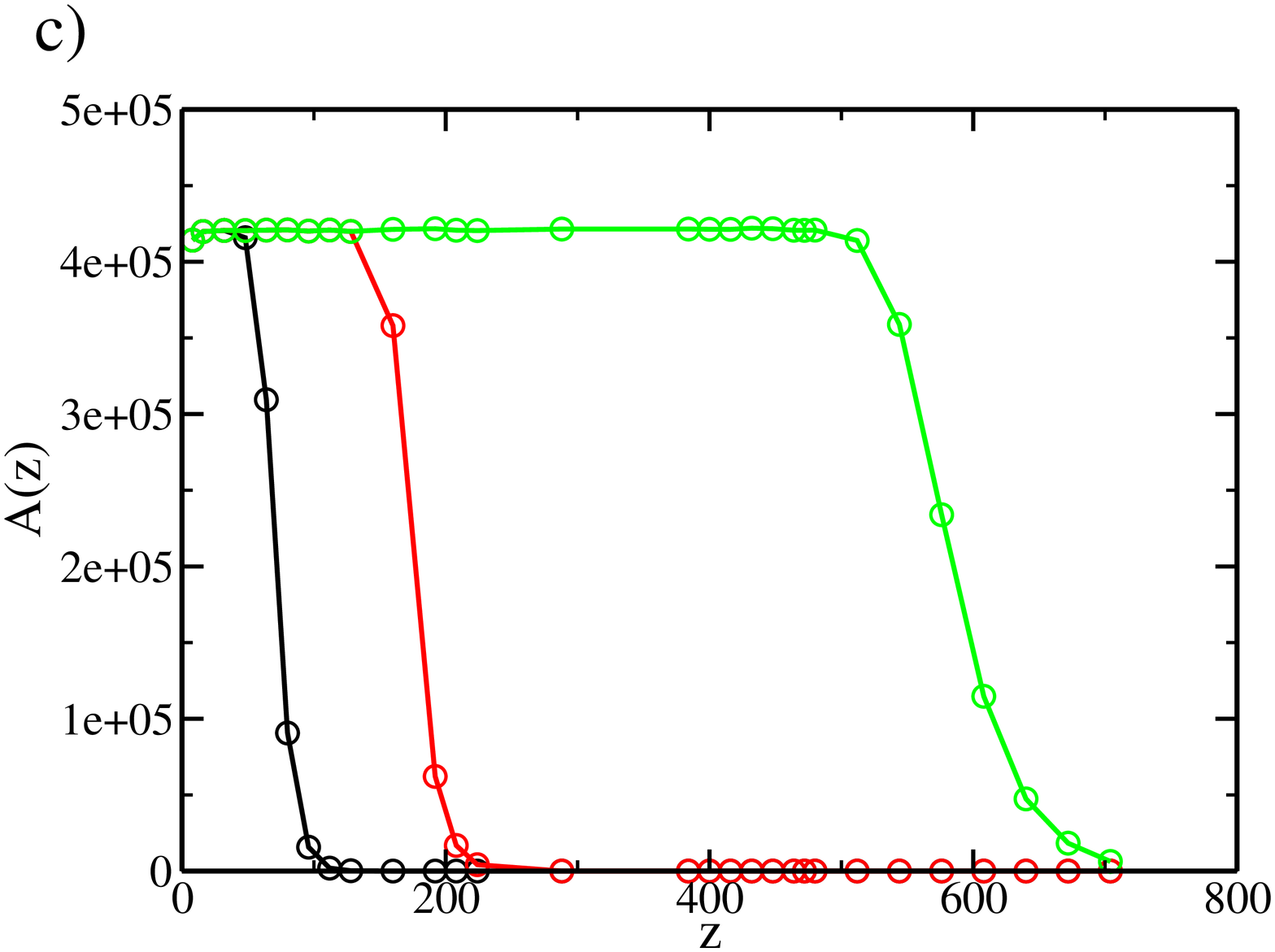}\\[2ex]
\includegraphics[width=6cm]{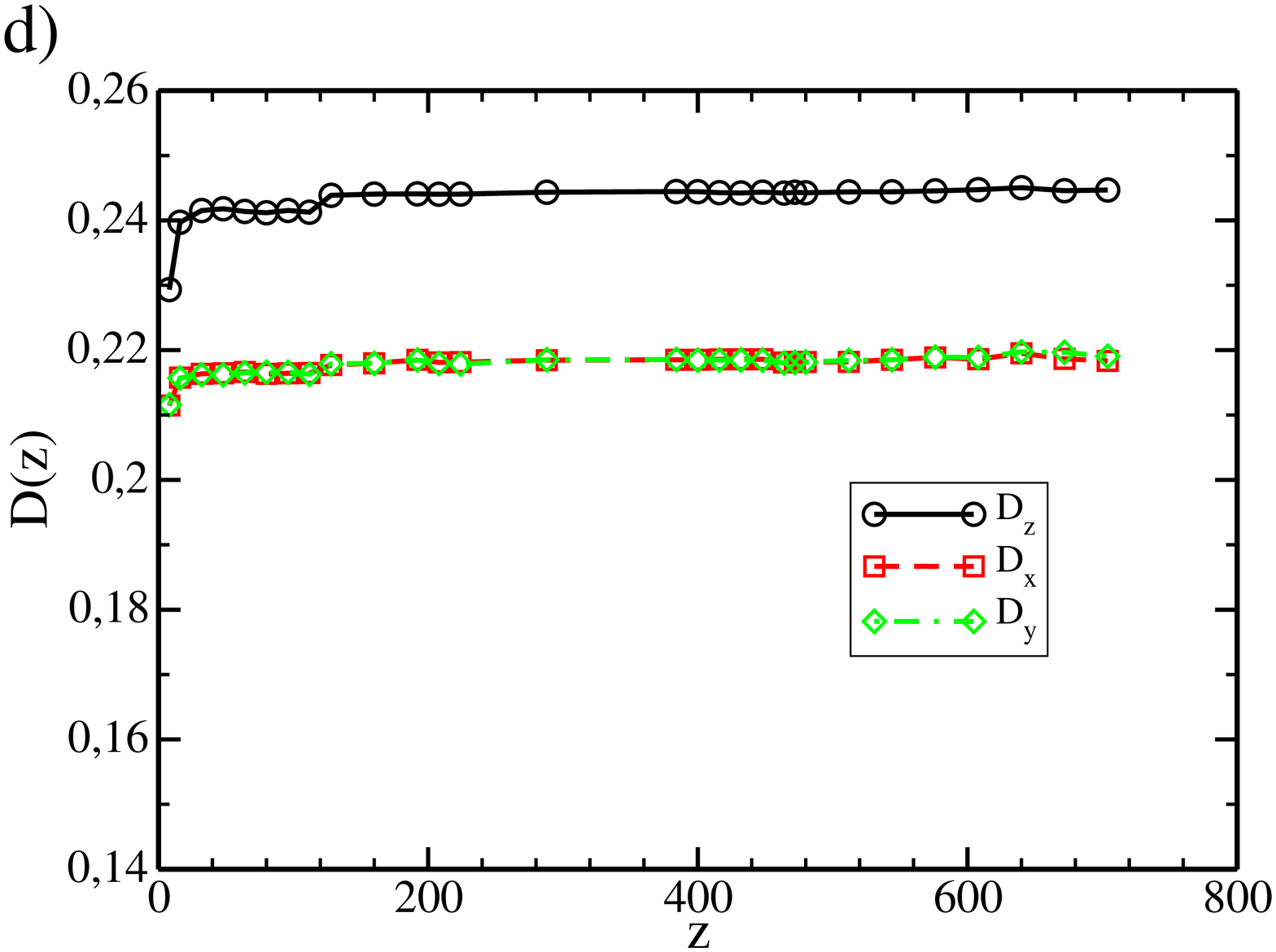}
\caption{(Color online)
Concentration profiles without (a) and with (b) time scaling, number of sites explored by RW (c), and
local diffusion coefficients (d) of tracer particles as a function of the depth $z$
in porous solids produced by BDNNN with rule (II) for the random steps of the RW.
Points and lines are defined in the caption to Fig. \ref{fig3}.
}
\label{fig6}
\end{figure}

Linear fits of the data for large $z$ are also shown in Figs. 3a, 4a, 5a, and 6a.
Their slopes are estimates of the exponent $\mu$ in Eq. \ref{rho}.
At short times ($t={10}^3$), the slopes are significantly below $2$ in all cases.
This is usually a transient behavior, since
the fits of the data at longer times ($t={10}^4$ and $t={10}^5$) in Figs. 3a, 4a, and 6a
have slopes close to the normal diffusion value $\mu =2$ (deviations are smaller than $10\%$).
However, in BDNNN deposits, the stretched concentration decay (slower than Gaussian)
is observed up to very long times, with fits in Fig. 5a ranging from $\mu\approx 1.33$ to
$\mu\approx 1.40$, in depths ranging from $z\sim 1$ to $z\approx 250$ (in units of the
lattice constant).

The plots of the concentration profiles are rescaled by the time $t$ in Figs. 3b, 4b, 5b, and 6b.
Reasonable data collapse of the large $z$ data is obtained in Figs. 3b, 4b, and 6b, using the
scaling variable $z/t^{1/2}$, which gives $\alpha/\mu=1/2$
[Eqs. (\ref{rho}) and (\ref{R})]. This is consistent with asymptotic normal diffusion.

On the other hand, walkers of type (I) in BDNNN deposits show much slower time
dependence of the concentration profile, which could be antecipated by comparison of Fig.~5a
with Figs. 3a, 4a, or 6a. A reasonable data collapse for large $z$ is obtained in Fig. 5b with
the scaling variable $z/t^{0.15}$, which gives $\alpha/\mu \approx 0.15$ [Eqs. (\ref{rho}) and (\ref{R})].
This estimate  gives $\langle z^2\rangle\sim t^{0.3}$, which indicates subdiffusion.

A comparison with experimental results is interesting at this point.

For $Pt$ atoms entering porous carbon samples, Ref. \protect\cite{brault1} gives $\mu\approx0.55$,
which characterizes a stretched decay (much slower than Gaussian) of the concentration profile, and
gives a very slowly increasing $R(t)$, with $\alpha\approx 0.2$. Combination of these results
gives $\langle z^2\rangle\sim t^{0.72}$, i. e. $\alpha/\mu=0.36$ [Eqs. (\ref{rho}) and (\ref{R})].
In the usual classification \cite{bouchaud,havlin}, a system with $\alpha/\mu<1/2$ is a case of subdiffusion
(however, the remarkable stretching of the concentration profile is a nontrivial feature that led
the authors of Ref. \protect\cite{brault1} to propose it was superdiffusive scaling).

For water diffusion in disordered colloidal systems, Ref. \protect\cite{palombo2011}
shows that several combinations of $\alpha<1$ and $\mu<2$ are
possible, which may give sub or superdiffusion. However, in that case, all estimates
are close to  $\alpha=1$ and $\mu=2$.

A recent work on $Pt$ atoms entering pores of anodic alumina \cite{brault3} reported
$\mu\approx 1/3$ and $\alpha\approx 1.25$, which gives a remarkably rapid
superdiffusion [$2\alpha /\mu\approx 7.5$; Eq. (\ref{z2})].
However, the organized geometry of those samples make them differ markedly from our deposits.

Finally, we recall that subdiffusion was also observed in porous deposits produced
by other ballistic-like models in Ref. \protect\cite{giri}.
Those deposits seem to have lower porosity than ours and have different pore shapes and connectivity. 
Only the mean-square displacement of the walkers was studied in that work, thus the present
results strongly suggest the study of concentration profiles in those samples.

\subsection{Explored area and diffusion coefficients}
\label{areacoef}

The anomalous diffusion for walkers of type (I) in BDNNN deposits is related to several
constraints for their steps, similarly to other systems \cite{havlin}.
First, the volume accessible for the walkers is restricted because the porosity is very
large (Fig. 1d) and they are always adsorbed to the internal solid walls. Second, since
the steps are restricted to NN (Fig. 2a), some movements become impossible, such as the
corner rounding shown for the upper walker in Fig. 2b. This is certainly an important
contribution to anomalous scaling.

Here we measure some quantities that help to explain the results presented
in Sec. \ref{concentration} and motivate the models presented in Sec. \ref{tube}.

The pore region accessible for the walkers is characterized by
the total area $A(z)$ explored by the walkers at each depth $z$, for several times.
This area is the total number of pore sites at depth $z$ that have been occupied by a walker at least once
up to time $t$. The explored areas $A(z)$ for three different times are shown in
Figs. 3c, 4c, 5c, and 6c, for the respective solids and step conditions.

The three cases with transient anomalous scaling (Figs. 3, 4, and 6) show similar
behavior of $A(z)$ for short and long times. For short times ($t\lesssim 1000$), there is a rapid
decay of $A(z)$ at depths varying from $\approx 40$ to $\approx 100$, depending on the
type of solid and the allowed steps.
The corresponding fits of the concentration profiles extend to larger depths and show anomalous 
scaling ($\mu<2$). For long times, the flat
region of $A(z)$ extends to much larger depths, which means that the walkers explore an
approximately constant cross section as they penetrate in the porous deposits.
The corresponding concentration profiles show normal scaling $\mu=2$.

For walkers of type (I) in BDNNN deposits, Fig. 5c shows a slow decay of $A(z)$ until very long times.
A fit of the data in Fig. 5c for $t={10}^5$ suggests an approximate decay as $A(z)\sim 1/z$.
Thus, there is an actual reduction of the area that can be reached by the walkers as they penetrate
in the porous solid. The porosity is approximately depth-independent, which means that a
large part of the porous space is not accessible to particles moving with the constraint of adsorption
to the internal walls.

Figs. 7a-d show cross sections of a BDNNN solid at four depths $z$ and highlights the
area occupied by walkers of type (I) at $t={10}^5$. It confirms the reduction of the
explored area with the depth. As will be shown in Sec. \ref{tube}, the stretched
decays of concentration profiles is intimately related to this feature.

\begin{figure}
\includegraphics[width=4cm]{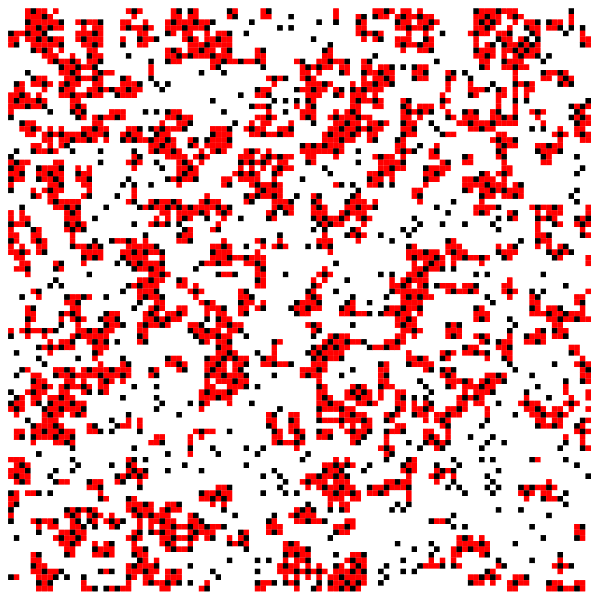}\\[2ex]
\includegraphics[width=4cm]{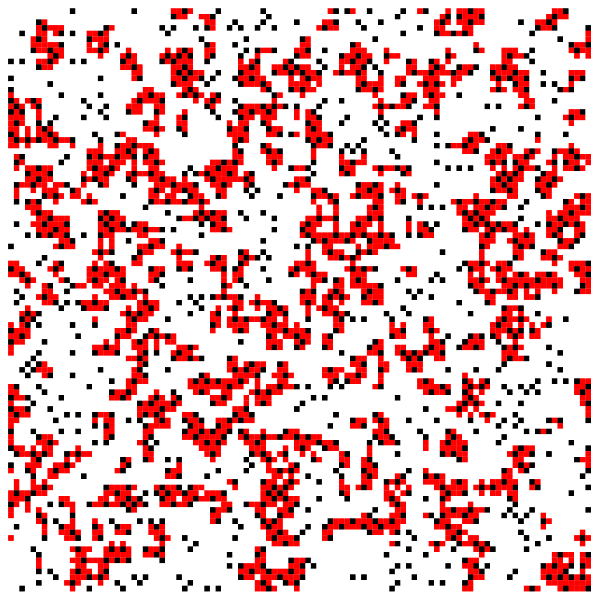}\\[2ex]
\includegraphics[width=4cm]{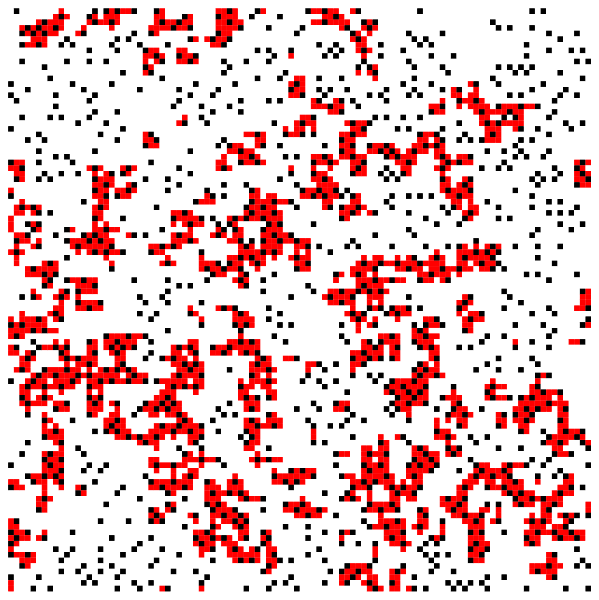}\\[2ex]
\includegraphics[width=4cm]{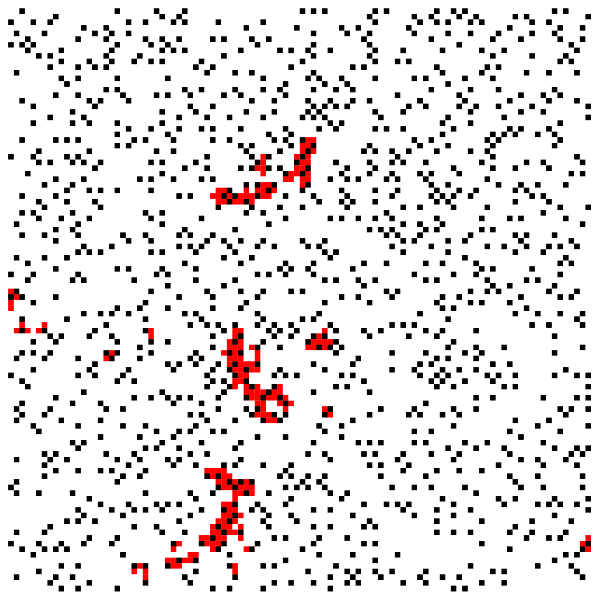}
\caption{(Color online) Cross sections of a porous solid produced by BDNNN at depths
(a) $z=16$, (b) $32$, (c) $96$, and (d) $128$, with solid part in black,
porous sites visited by RW [rule (I)] at $t={10}^5$ in red, and non-visited porous
part in white.}
\label{fig7}
\end{figure}

Many works on anomalous diffusion connect this feature to
position-dependent diffusion coefficients
(see e. g. Refs. \protect\cite{malacarne,pedron}). In the present systems, this possibility is
ruled out by measuring local diffusion coefficients that characterize the diffusion process
in a narrow range around each depth $z$ (in opposition to asymptotic coefficients that would
characterize the diffusion in the full pore network).

The local diffusion coefficient at height $z$ is measured by tracers left at all accessible points
in the corresponding depth of the solid. The number of tracers left at each point $(x,y,z)$
is proportional
to the number of walkers which have occupied that position in the course of the
original simulations up to $t={10}^5$ (i. e. in the simulations with walkers left at the
external surface).
Each tracer tries to execute a small number of steps, $t_0=100$, following the same rules
[(I) or (II)] of the RW. 
For the set of tracers starting at a given depth $z'$, the mean-square vertical displacement
$\langle {\left( \Delta z\right)}^2\rangle\left( z',t_0\right)$ is measured after $t_0$ steps
(averaging over $x$, $y$, and different tracers).
The resulting local diffusion coefficient at depth $z'$ is estimated as
\begin{equation}
D_z\left( z'\right) = \frac{\langle {\left(\Delta z\right)}^2\rangle\left( z',t_0\right)}{t_0} .
\label{difcoef}
\end{equation}
$D_x\left( z'\right)$ and $D_y\left( z'\right)$ are equivalently measured.

Figs. 3d, 4d, 5d, and 6d show $D_z\left( z\right)$ as a function of the depth $z$,
for the respective solids and walker
types of Figs. 3-6. Considering the small values of the diffusion coefficients, we observe
that only a small region around the original position is explored, thus
we are actually measuring local diffusion coefficients.
In all cases, the coefficients are approximately depth-independent,
which means that the porous solids are homogeneous at short lengthscales. Thus, the anomalous
diffusion observed in these systems cannot be explained by changes in the local diffusion coefficient.
This contrasts to some widely used models of anomalous diffusion \cite{malacarne,pedron} and
the proposed generalized diffusion equation of Ref. \protect\cite{brault1}.

Figs. 3d, 4d, 5d, and 6d also show the diffusion coefficients $D_x(z)$ and $D_y(z)$, with
no significant dependence on $z$. The estimates of $D_x(z)$ and $D_y(z)$ are nearly the same, which
confirms the homogeneity in the horizontal directions and the accuracy of the simulations.

We conclude that local features of the porous deposits are not responsible for the anomalous diffusion
observed here. Instead, the anomaly is related to the different accessibility of
inner sites at a given height, since they have different connections (if any) to the outer surface.

\section{Diffusion in tubes of variable cross section}
\label{tube}

The work with RW entering porous deposits showed anomalous diffusion when
the available area for the walkers decreased with the depth, but with constant diffusion
coefficients. For this reason, this second part of the paper
considers a simple model with those features.

\subsection{Basic equations and crossover}
\label{crossover}

We consider an ensemble of RW starting at randomly chosen positions with $z=0$
at $t=0$, confined to move in the three-dimensional region $z\geq 0$,
$x^2+y^2\geq {\left[ A\left( z\right)/\pi\right]}$,
which defines a tube of $z$-dependent cross section $A(z)$. Here we are interested in cases
where $A(z)$ is monotonically decreasing, similarly to the porous media
produced by BDNNN. An illustration is provided in Fig. 8a.

\begin{figure}
\includegraphics[width=8.5cm]{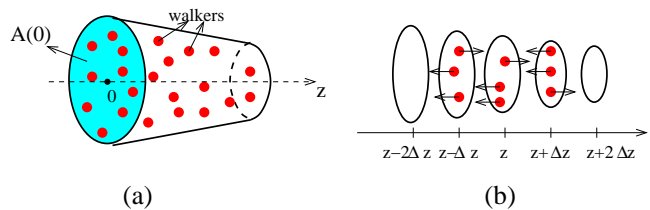}
\caption{(Color online) (a) Tube with $z$-dependent cross section and walkers
moving inside it; (b) Random steps of the walkers in the tube, with larger
number of steps towards the larger neighboring areas.
}
\label{fig8}
\end{figure}

Let $N(z,t)$ be the number of walkers at position $z$ at time $t$. This quantity is
proportional to the walker concentration $\rho (z,t)$ if the latter is measured over a
constant cross section, i. e. the cross section of a solid containing the empty tube.
This is similar to the measurement of concentration in the porous deposits
of Sec. \ref{concentration} and in experimental works, since the total number of
walkers at each depth $z$ is divided by the (constant) area of a slice that includes solid
parts and pores.

In the limit of continuous $z$ and $t$, the change in $N(z,t)$ in a small time interval is
\begin{eqnarray}
&& N\left( z,t+\Delta t\right) - N\left( z,t\right) = \nonumber\\
&& -k N\left( z,t\right) +
k N\left( z-\Delta z,t\right) \frac{A\left( z\right)}{A\left( z-2\Delta z\right)+A\left( z\right)}
+\nonumber\\
&& k N\left( z+\Delta z,t\right) \frac{A\left( z\right)}{A\left( z+2\Delta z\right)+A\left( z\right)} ,
\label{Nzt}
\end{eqnarray}
where $k$ is a constant (increasing with $\Delta t$ and decreasing with $\Delta z$).
Eq. (\ref{Nzt}) accounts for the fact that all neighboring sites are equally probable
for a random step, thus the number of walkers that move from $z\pm \Delta z$ to a neighboring
position ($z\pm 2\Delta z$ or $z$) is proportional to the area available at that position.
This is illustrated in Fig. 8b, with a larger number of walkers moving to the larger neighboring area.

Considering that $A\left( z\pm 2\Delta z\right)+A\left( z\right)\approx 2 A\left( z\pm\Delta z\right)$
and dividing Eq. (\ref{Nzt}) by $A\left( z\right)$, we obtain an equation for
$N\left( z,t\right)/A\left( z\right)$. Defining $D\equiv k{\Delta z}^2/\Delta t$ and 
taking the continuum limit ($\Delta z\to 0$, $\Delta t\to 0$), we obtain
\begin{equation}
\frac{\partial \left( N/A\right) }{\partial t} = \frac{D}{2} \frac{\partial^2 \left( N/A\right)}{\partial z^2} .
\label{diffeq}
\end{equation}
This is the usual diffusion equation for the density $N(z,t)/A(z)$.

The solution of Eq. (\ref{diffeq}) is 
\begin{equation}
N\left( z,t\right)= N_0 A\left( z\right) \exp{\left[ -z^2/\left( 2Dt\right)\right]} ,
\label{Nsol}
\end{equation}
where $N_0$ is a normalization constant dependent on $t$ (typically as a power law, which is
slowly varying compared to the exponential factor); for instance,
for constant $A$, the normal value $N_0A=\displaystyle\frac{1}{\sqrt{2Dt}}$ is recovered.
This solution for $N\left( z,t\right)$ is a normal diffusion behavior with the concentration
at position $z$ templated by the area of the tube.

Rewriting the walker concentration of Eq. (\ref{Nsol}) as
\begin{equation}
N \sim \exp{\left[ -z^2/\left( 2Dt\right) + \ln{A\left( z\right)}\right]} ,
\label{Nsolcross}
\end{equation}
a crossover is expected if $|\ln{A\left( z\right)}|$ does not scale as $z^2$, i. e.
non-Gaussian $A\left( z\right)$. 
Here we restrict the discussion to the cases in which $A\left( z\right)$ decreases slower than
a Gaussian.
In these cases, the first term in the exponential of
Eq. (\ref{Nsolcross}) is dominant for large $z$, but the second one is larger for
small $z$, particularly for long times.
A crossover position $z_c$ is found by matching the two terms in the exponential of
Eq. (\ref{Nsolcross}), which gives
\begin{equation}
{z_c}^2/\left( 2Dt\right) \approx \ln{A\left( z_c\right)} .
\label{zcr}
\end{equation}
Normal diffusion concentration is observed for $z\gg z_c$ and anomalous
diffusion is observed for $z\ll z_c$, with a stretched exponential decay (slower than Gaussian)
of the concentration profile.

As time increases, $z_c$ increases, so
the anomalous diffusion regime extends to larger regions and to lower densities. 
In this regime, matching the dominant
terms of Eqs. (\ref{rho}) and (\ref{Nsol}), the anomaly exponent is given by power-counting in the relation
$\ln{A\left( z\right)}\sim z^{\mu}$.

Normal diffusion is able to distribute a large particle density in domains
of size
\begin{equation}
z_{diff}\sim \sqrt{Dt} ,
\label{zdiff}
\end{equation}
thus this is a region far from the Gaussian tail of the
concentration profile. If diffusion takes place in a physically confined region, then
the particle density (taken over porous and solid regions) is templated by the shape of the
confined pore space up to $z\approx z_{diff}$. In other words, the concentration profile scales
similarly to $A(z)$ in this highly populated region.
Consequently, it is the templating effect of the tube shape that leads to a concentration decay
similar to $A(z)$ for small $z$ and the crossover phenomenon described above.

\subsection{Tubes with $A(z)\sim A_0/z$}
\label{A1overz}

As a first numerical application, we consider a tube with $A(z)= A_0 / z$, where $A_0=10^4$. This is a decay
similar to the explored area of RW of type (I) in BDNNN deposits (Sec. \ref{porous}).
Simulations of ${10}^8$ different walks were performed. In each walk, the particle executes one random step to
a NN site per time unit, thus $D=1/3$ in Eq. (\ref{diffeq}). 
Eq. (\ref{Nsol}) gives $N\sim \displaystyle\frac{A_0}{z} \exp{\left[ -z^2/\left( 2Dt\right)\right]}$.

In Fig. 9, the usual scaling plot of the concentration profile is shown.
The scaling variable $z/t^{0.43}$ provides the best fit for large $z$, where the
slope is near $0.7$, which is much smaller than the normal diffusion exponent
$\mu=2$. That estimate is obtained in a narrow region of the
scaling variable, thus it may not be viewed as a true scaling exponent.
However, taking it as an effective estimate of $\mu$ and taking $\alpha\approx 0.43$
(from the scaling variable in Fig. 9),
we obtain $\langle z^2\rangle \sim t^{0.86}$ [Eqs. (\ref{rho}) and (\ref{R})].
This corresponds to an apparent subdiffusion.

\begin{figure}
\includegraphics[width=7cm]{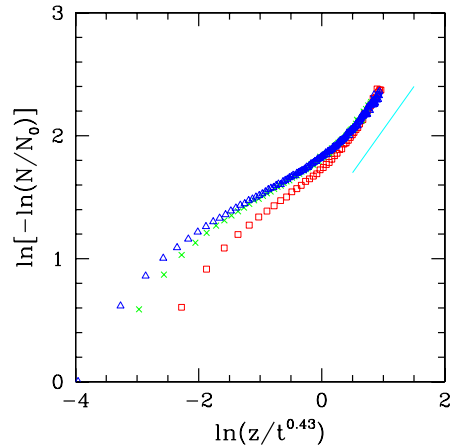}
\caption{(Color online)
Normalized walker concentration as a function of the scaled depth in a tube with
$A(z)= A_0/ z$ at $t=1000$ (red squares), $t=5000$ (green crosses), and
$t=10000$ (blue triangles) and the solid line has slope $0.7$.
}
\label{fig9}
\end{figure}

These results and those for walkers of type (I) in BDNNN deposits show important similarities.
Both systems have stretched exponential decay of the concentration profile, a feature that
is enhanced in the tube. They also show slowly increasing (subdiffusive) mean-square displacement,
this effect being enhanced in the porous deposits.
The slopes of scaling plots differ due to the different microscopic structures.

Surprisingly, the effective exponents obtained from Fig. 9 ($\mu\approx 0.7$, $\alpha\approx 0.3$)
are much closer to the ones for $Pt$ atoms entering porous carbon samples
($\mu\approx 0.55$ and $\alpha\approx 0.2$) \cite{brault1}, with the particularly interesting
feature of highly stretched concentration profile.
This suggests that the anomalous scaling observed in $Pt$ atom diffusion may be a consequence of a
very long crossover due to the confined geometry of those samples.

\subsection{Tubes with exponentially decreasing cross section}
\label{exponential}

Here we  consider the case of exponential decay of $A\left( z\right)$
because this facilitates the analytical study of crossover effects:
\begin{equation}
A\left( z\right)=A_0\exp{\left( -z/z_0\right)} .
\label{Aexp}
\end{equation}
This gives $N\left( z,t\right)=N_1\exp{\left[ -z^2/\left( 2Dt\right) -z/z_0\right]}$,
where $N_1$ depends on $A_0$, $z_0$, and $t$ [$N_1$ should not be confused with
$N_0$ in Eq. (\ref{Nsol}): $N_0=N_1 \exp{\left( -z/z_0\right)}$].
The crossover position is
\begin{equation}
z_c \approx 2Dt/z_0 
\label{zcrexp}
\end{equation}
and the concentration relative to the origin at the crossover region is
\begin{equation}
\rho_c\equiv N\left( z_c,t\right)/N\left( 0,t\right) \sim \exp{\left( -4Dt/{z_0}^2\right)}.
\label{rhocrexp}
\end{equation}

For $z\gg z_c$, the asymptotically normal diffusion is observed, with density $\rho\ll \rho_c$.
This corresponds to very small concentrations at long times, thus this regime is difficult
to be observed in simulation or in possible experimental realizations.
For $z\ll z_c$, $N\left( z,t\right)$ follows the simple exponential decay of $A(z)$,
which is the templating effect of the tube shape.
The diffusion seems to be anomalous, with $\mu\approx 1$, and the
concentration is much larger than $\rho_c$.
As time increases, this regime extends to larger distances and lower densities.

The mean-square displacement
$\langle z^2\left( t\right)\rangle\sim \int_{0}^{\infty}{z^2\rho\left( z,t\right) dz}$
is estimated by finding the maximal value of $z^2\rho\left( z,t\right)$, which occurs for
$z=z_{max}=\left[\sqrt{Dt/{z_0}^2+8}-\sqrt{Dt}/z_0\right]\sqrt{Dt}/2$. For $z_0\gg \sqrt{Dt}\sim z_{diff}$
[Eq. (\ref{zdiff})], normal scaling with  $\langle z^2\rangle\sim t$ is obtained. This is a case in which
the tube does not template the concentration profile. For $z_0\ll \sqrt{Dt}\sim z_{diff}$,
$\langle z^2\rangle\approx {z_0}^2$, with corrections in $1/t$, since
the concentration profile follows the tube shape [$\rho\sim A\left( z\right)$ in the region
contributing to $\langle z^2\rangle$]. This is a case of anomalously slow diffusion ($\alpha =0$).
These limiting behaviors indicate that subdiffusion is expected in the crossover region.

The crossover is illustrated in Fig. 10a for a tube with $z_0=20$ at three different times.
The distributions were obtained in simulations of ${10}^7$ different walks with $D=1/3$.
For $t={10}^3$, a crossover is observed from an initial regime with slope $\approx 0.6$ to
a final regime with slope near $2$. Again, we stress that these values are
effective exponents representative only of a narrow scaling region.
Eq. (\ref{zcrexp}) predicts the crossover length
$z_c\approx 33$ for $z_0=20$ and $t={10}^3$, in good agreement with the results in Fig. 10a.
For $t={10}^5$, Fig. 10a does not show any signature of a crossover up to $z\approx 180$,
which is the maximal depth in which accurate concentrations could be estimated.
Moreover, the data in Fig. 10a do not fit a single scaling curve with any
variable of the form $z/t^{\alpha/\mu}$.

\begin{figure}
\includegraphics[width=7cm]{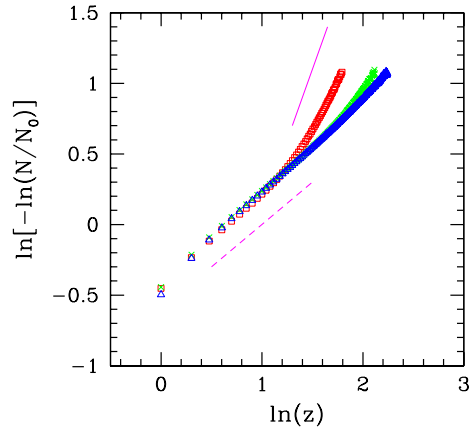}\\[2ex]
\includegraphics[width=7cm]{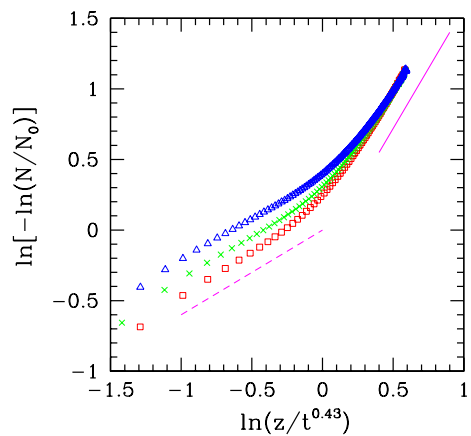}
\caption{(Color online) (a) Normalized walker concentration as a function of the depth in a tube with
exponentially decreasing $A(z)$ with $z_0=20$ at $t=1000$ (red squares), $t=10000$ (green crosses),
and $t=100000$ (blue triangles). The dashed line has slope $0.6$ and the solid line has slope $2$.
(b) Normalized walker concentration as a function of the scaled depth in a tube with
exponentially decreasing $A(z)$ with $z_0=50$ at $t=1000$ (red squares), $t=2000$ (green crosses),
and $t=5000$ (blue triangles). The dashed line has slope $0.6$ and the solid line has slope $1.7$.
}
\label{fig10}
\end{figure}

An illustration of the crossover scaling is possible in tubes with $z_0=50$
and a smaller time range.
Fig. 10b shows $\ln{\left[ -\ln{(N/N_0)}\right]}$ as a function of $\ln{z/t^{0.43}}$
at $t=1000$, $t=5000$, and $t=10000$. The traditional form of the scaling variable gives a
reasonable data
collapse for large $z$ in this system and indicates an effective anomalous diffusion
(note that a complete
scaling plot in this problem would have to involve the crossover length $z_c$).
The initial slope is near $0.6$ and crosses over to a slope near $1.7$
for the largest values of~$z$. These values are close to the predictions of effective
exponents $1$ and $2$ of the crossover scaling;
deviations are justified by the limited range of $z$ in which the concentration is not negligible. 
The crossover positions predicted by Eq. (\ref{zcr}) range from $z_c\approx 13$ ($t=1000$) to
$z_c\approx 130$ ($t=10000$), which are close to the points with largest curvatures in the plots
of Fig. 10b.

\section{Conclusion}
\label{conclusion}

In the first part of this work, we studied the statistics of RW starting at the outer
surface of porous deposits and restricted to move along their internal
walls. BD and BDNNN models were used to produce the porous solids.
These conditions are chosen to parallel the penetration of metal atoms
in porous materials during the deposition of a film at their external surfaces.
The concentration decay with the depth $z$ is slower than Gaussian at short times.
At long times, a change to normal diffusion is observed in
most cases. However, the stretched exponential concentration profile remains in the most
porous solids, produced by BDNNN, if the walker steps are restricted to nearest neighbors.
This feature is correlated to a decay of the area explored by the RW approximately with the inverse $z$.
However, RW of tracer particles left at various points of the solid do not show a significant
dependence of the diffusion coefficient with $z$. This rules out the description of
the anomalous diffusion by equations with position-dependent coefficients
\cite{malacarne,pedron}.
This shows that local features in space or time are not of main importance for the anomaly,
but the global history in the penetration of the diffusing particles is the main ingredient. 
Starting from surface deposition sites, not all inner sites at a given height are
accessible in an equivalent way,
some sites having a preferential path connecting them to the outer surface.  

In the second part of the work, we propose a model of RW confined to a tube of decreasing cross
section $A(z)$.
We first consider an area which decreases slower than a Gaussian, similar to the decay
obtained in one of the simulated experiment of the first part and account for a subdiffusion
regime. Then considering the specific case of $A(z)$ with a simple exponential decay,
we predict analytically the crossover from a subdiffusion to a normal diffusion regime.
The crossover position and crossover density show good agreement with numerical data.
In both cases, the anomalous scaling is understood as a
templating effect of $A(z)$ on the concentration profile, which is measured in solid slices of constant
area, but with walkers exploring only a fraction of this area.

The above results suggest an alternative explanation to the stretched exponential concentration profiles
and anomalous diffusion shown in recent works on the penetration of $Pt$ atoms in disordered porous
deposits \cite{brault1}, despite the difference in the exponent values.
An interesting feature of our models is the independence of diffusion coefficients on the position,
which is related to the local homogeneity of the media, while the model equations used in Ref.
\protect\cite{brault1,brault3} assume a dependence of those coefficients in time and position. 
On the other hand, those experiments also include a flux of atoms to the outer surface of the
porous media, which is a feature not included in the present models and certainly important for their
quantitative description.

Independently of this possible application, the present work highlights the difficulty in  interpreting
experimental or computational data on anomalous diffusion, particularly if only the concentration profiles
are measured or large crossover times are present.
This is in agreement with recent experimental and theoretical works on the subject
\cite{saxton,raccis,korabel,kennedy}.


\begin{acknowledgments}
This work was supported by CNPq and FAPERJ (Brazilian agencies).
\end{acknowledgments}



\vfill\eject

\end{document}